\documentclass[10pt]{article}

\usepackage{amsmath}
\usepackage{amssymb}

\usepackage{graphicx}

\usepackage{cite}

\usepackage{color} 

\usepackage{setspace} 
\usepackage[labelfont=bf,labelsep=period,justification=raggedright]{caption}

\makeatletter
\renewcommand{\@biblabel}[1]{\quad#1.}
\makeatother

\date{}

\begin{document}

\begin{flushleft}
{\Large
\textbf{A Self-Organizing State-Space-Model Approach for Parameter Estimation in Hodgkin-Huxley-Type Models of Single Neurons}
}
\\
Dimitrios V. Vavoulis$^{1,\ast}$, 
Volko A. Straub$^{2}$, 
John A. D. Aston$^{3}$,
Jianfeng Feng$^{1,*}$
\\
\bf{1} Dept. of Computer Science, University of Warwick, Coventry, UK
\\
\bf{2} Dept. of Cell Physiology and Pharmacology, University of Leicester, Leicester, UK
\\
\bf{3} Dept. of Statistics, University of Warwick, Coventry, UK
\\
$\ast$ E-mail: Dimitris.Vavoulis@dcs.warwick.ac.uk, Jianfeng.Feng@warwick.ac.uk
\end{flushleft}

\section*{Abstract}

Traditional approaches to the problem of parameter estimation in biophysical
models of neurons and neural networks usually adopt a global search
algorithm (for example, an evolutionary algorithm), often in combination
with a local search method (such as gradient descent) in order to
minimize the value of a cost function, which measures the discrepancy
between various features of the available experimental data and model
output. In this study, we approach the problem of parameter estimation
in conductance-based models of single neurons from a different perspective.
By adopting a hidden-dynamical-systems formalism, we expressed parameter
estimation as an inference problem in these systems, which can then
be tackled using a range of well-established statistical inference
methods. The particular method we used was Kitagawa's self-organizing
state-space model, which was applied on a number of Hodgkin-Huxley-type
models using simulated or actual electrophysiological data. We showed
that the algorithm can be used to estimate a large number of parameters,
including maximal conductances, reversal potentials, kinetics of ionic
currents, measurement and intrinsic noise, based on low-dimensional
experimental data and sufficiently informative priors in the form
of pre-defined constraints imposed on model parameters. The algorithm
remained operational even when very noisy experimental data were used.
Importantly, by combining the self-organizing state-space model with
an adaptive sampling algorithm akin to the Covariance Matrix Adaptation
Evolution Strategy, we achieved a significant reduction in the variance
of parameter estimates. The algorithm did not require the explicit
formulation of a cost function and it was straightforward to apply
on compartmental models and multiple data sets. Overall, the proposed
methodology is particularly suitable for resolving high-dimensional
inference problems based on noisy electrophysiological data and, therefore,
a potentially useful tool in the construction of biophysical neuron
models.

\section*{Author Summary}

Parameter estimation is a problem of central importance and, perhaps,
the most laborious task in biophysical modeling of neurons and neural
networks. An emerging trend is to treat parameter estimation in this
context as yet another statistical inference problem, which can be
tackled using well-established methods from Computational Statistics.
Inspired by these recent advances, we adopted a self-organizing state-space-model
approach augmented with an adaptive sampling algorithm akin to the
Covariance Matrix Adaptation Evolution Strategy in order to estimate
a large number of parameters in a number of Hodgkin-Huxley-type models
of single neurons. Parameter estimation was based on noisy electrophysiological
data and involved the maximal conductances, reversal potentials, levels
of noise and, unlike most mainstream work, the kinetics of ionic currents
in the examined models. Our main conclusion was that parameters in
complex, conductance-based neuron models can be inferred using the
aforementioned methodology, if sufficiently informative priors regarding
the unknown model parameters are available. Importantly, the use of
an adaptive algorithm for sampling new parameter vectors significantly
reduced the variance of parameter estimates. Flexibility and scalability
are additional advantages of the proposed method, which is particularly
suited to resolve high-dimensional inference problems.

\section*{Introduction}

Among several tools at the disposal of neuroscientists today, data-driven
computational models have come to hold an eminent position for studying
the electrical activity of single neurons and the significance of
this activity for the operation of neural circuits\cite{Herz:2006ys, De-Schutter:2005fk, Grillner:2003kx, Marder:1996vn}.
Typically, these models depend on a large number of parameters, such
as the maximal conductances and kinetics of gated ion channels. Estimating
appropriate values for these parameters based on the available experimental
data is an issue of central importance and, at the same time, the
most laborious task in single-neuron and circuit modeling.

Ideally, all unknown parameters in a model should be determined directly
from experimental data analysis. For example, based on a set of voltage-clamp
recordings, the type, kinetics and maximal conductances of the voltage-gated
ionic currents flowing through the cell membrane could be determined\cite{Willms:1999ly}
and, then, combined in a conductance-based model, which replicates
the activity of the biological neuron of interest under current-clamp
conditions with sufficient accuracy. Unfortunately, this is not always
possible, especially for complex compartmental models, which contain
a large number of ionic currents. 

A first problem arises from the fact that not all parameters can be
estimated within an acceptable error margin, especially for small
currents and large levels of noise. A second problem arises from the
practice of estimating different sets of parameters based on data
collected from different neurons of a particular type, instead of
estimating all unknown parameters using data collected from a single
neuron. Different neurons of the same type may have quite different
compositions of ionic currents\cite{Goldman:2001ve, Golowasch:2002qf, Prinz:2003nx, Prinz:2004cr}
(but, see also \cite{Nowotny:2007bh}). This implies that combining
ionic currents measured from different neurons in the same model or
even using the average of several parameters calculated over a population
of neurons of the same type will not necessarily result in a model
that expresses the experimentally recorded patterns of electrical
activity under current-clamp conditions. Usually, only some parameters
are well characterized, while others are difficult or impossible to
measure directly. Thus, most modeling studies rely on a mixture of
experimentally determined parameters and estimates of the remaining
unknown ones using automated optimization methodology (see, for example,
\cite{Lepora:2011fv, Hendrickson:2011hc, Keren:2009zr, Pospischil:2008bs, Hobbs:2008oq, Nowotny2008, Reid2007, Druckmann:2007ij, Achard:2006kl, Keren:2005tg, Tabak2000, Vanier:1999kl}).
Typically, these methods require the construction of a cost function
(for measuring the discrepancy between various features of the experimental
data and the output of the model) and an automated parameter selection
method, which iteratively generates new sets of parameters, such that
the value of the cost function progressively decreases during the
course of the simulation (see \cite{Van-Geit:2008dz} for a review).
Popular choices of such methods are evolutionary algorithms, simulated
annealing and gradient descent methods. Often, a global search method
(i.e. an evolutionary algorithm) is combined with local search (gradient
descent) for locating multiple minima of the cost function with high
precision. Since a poorly designed cost function (for example, one
that merely matches model and experimental membrane potential trajectories)
can seriously impede optimization, the construction of this function
often requires particular attention (see, for example, \cite{LeMasson2001}).
Nevertheless, these computationally intensive methodologies have gained
much popularity, particularly due to the availability of powerful
personal computers at consumer-level prices and the development of
specialized optimization software (e.g. \cite{Van-Geit:2007fu}). 

Alternative approaches also exist as, for example, methods based on
the concept of synchronization between model dynamics and experimental
data\cite{Tokuda2003}. An emerging trend in parameter estimation
methodologies for models in Computational Biology is to recast parameter
estimation as an inference problem in hidden dynamical systems and
then adopt standard Computational Statistics techniques to resolve
it\cite{Lillacci2010,Huys:2009qa}. For example, a particular study
following this approach makes use of Sequential Monte Carlo methods
(\textit{particle filters}) embedded in an Expectation Maximization
(EM) framework\cite{Huys:2009qa}. Given a set of electrophysiological
recordings and a set of dynamic equations that govern the evolution
of the hidden states, at each iteration of the algorithm the expected
joint log-likelihood of the hidden states and the data is approximated
using particle filters (Expectation Step). At a second stage during
each iteration (Maximization Step), the log-likelihood is locally
maximized with respect to the unknown parameters. The advantage of
these methods, beyond the fact that they recast the estimation problem
in a well-established statistical framework, is that they can handle
various types of noisy biophysical data made available by recent advances
in voltage and calcium imaging techniques.

Inspired by this emerging approach, we present a method for estimating
a large number of parameters in Hodgkin-Huxley-type models of single
neurons. The method is a version of Kitagawa's self-organizing state-space
model\cite{Kitagawa1998} combined with an adaptive algorithm for
selecting new sets of model parameters. The adaptive algorithm we
have used is akin to the Covariance Matrix Adaption (CMA) Evolution
Strategy\cite{Igel:2007}, but other methods (e.g. Differential Evolution
as described in \cite{Price:2005}) may be used instead. We demonstrate
the applicability of the algorithm on a range of models using simulated
or actual electrophysiological data. We show that the algorithm can
be used successfully with very noisy data and it is straightforward
to apply on compartmental models and multiple datasets. An interesting
result from this study is that by using the self-organizing state-space
model in combination with a CMA-like algorithm, we managed to achieve
a dramatic reduction in the variance of the inferred parameter values.
Our main conclusion is that a large number of parameters in a conductance-based
model of a neuron (including maximal conductances, reversal potentials
and kinetics of gated ionic currents) can be inferred from low-dimensional
experimental data (typically, a single or a few recordings of membrane
potential activity) using the algorithm, if sufficiently informative
priors are available, for example in the form of well-defined ranges
of valid parameter values.

\section*{Methods}

\subsection*{Modeling Framework}

We begin by presenting the current conservation equation that describes
the time evolution of the membrane potential for a single-compartment
model neuron: 
\begin{equation}
\frac{dV}{dt}=\frac{I_{ext}-G_{L}(V-E_{L})-\sum_{i}I_{i}}{C_{m}}\label{eq:dVdt}
\end{equation}
where $V$, $I_{ext}$ and $I_{i}$ are all functions of time. In
the above equation, $C_{m}$ is the membrane capacitance, $V$ is
the membrane potential, $I_{ext}$ is the externally applied (injected)
current, $G_{L}$ and $E_{L}$ are the maximal conductance and reversal
potential of the leakage current, respectively, and $I_{i}$ is the
$i^{th}$ transmembrane ionic current. A voltage-gated current $I_{i}$
can be modeled according to the Hodgkin-Huxley formalism, as follows:
\begin{equation}
I_{i}=G_{i}m_{i}^{p_{i}}h_{i}(V-E_{i})
\end{equation}
where $m_{i}$ and $h_{i}$ are both functions of time. In the above
expression, $G_{i}$ and $E_{i}$ are the maximal conductance and
reversal potential of the $i^{th}$ ionic current, $m_{i}$ and $h_{i}$
are dynamic gating variables, which model the voltage-dependent activation
and inactivation of the current, and $p_{i}$ is a small positive
integer power (usually, not taking values larger than 4). The product
$m_{i}^{p_{i}}h_{i}$ is the proportion of open channels in the membrane
that carry the $i^{th}$ current. The gating variables $m_{i}$ and
$h_{i}$ obey first-order relaxation kinetics, as shown below: 

\begin{equation}
\frac{dm_{i}}{dt}=\frac{m_{\infty,i}-m_{i}}{\tau_{m_{i}}}\qquad,\qquad\frac{dh_{i}}{dt}=\frac{h_{\infty,i}-h_{i}}{\tau_{h_{i}}}\label{eq:dxdt}
\end{equation}
where the steady states ($m_{\infty,i}$ , $h_{\infty,i}$) and relaxation
times ($\tau_{m_{i}}$ , $\tau_{h_{i}}$) are all functions of voltage. 

Using vector notation, we can write the above system of Ordinary Differential
Equations (ODEs) in more concise form: 

\begin{equation}
\frac{d\mathbf{x}(t)}{dt}=\mathbf{f}(\mathbf{x}(t),t)\label{eq:vec}
\end{equation}
where the state vector $\mathbf{x}(t)$ is composed of the time-evolving
state variables \textbf{$V$}, $m_{i}$ and $h_{i}$ and the vector-valued
function $\mathbf{f}(\cdot,\cdot)$, which describes the evolution
of $\mathbf{x}(t)$ in time, is formed by the right-hand sides of
Eqs. \ref{eq:dVdt} and \ref{eq:dxdt}. Notice that $\mathbf{f}(\cdot,\cdot)$
also depends on a parameter vector $\theta$, which for now is dropped
from Eq. \ref{eq:vec} for notational clarity. Components of $\theta$
are the maximal conductances $G_{i}$, the reversal potentials $E_{i}$
and the various parameters that control the voltage-dependence of
the steady states and relaxation times in Eq. \ref{eq:dxdt}. 

The above deterministic model does not capture the inherent variability
in the electrical activity of neurons, but rather some average behavior
of intrinsically stochastic events. In general, this variability originates
from various sources, such as the random opening and shutting of transmembrane
ion channels or the random bombardment of the neuron with external
(e.g. synaptic) stimuli\cite{Faisal:2008fk}. Here, we model the inherent
variability in single-neuron activity by augmenting Eq. \ref{eq:vec}
with a noisy term and re-writing as follows:

\begin{equation}
d\mathbf{x}(t)=\mathbf{f}(\mathbf{x}(t),t)dt+\sqrt{\Sigma_{x}}d\mathbf{W}_{\mathbf{x}}(t)\label{eq:stoch}
\end{equation}
where $\Sigma_{\mathbf{x}}$ is a covariance matrix and $\mathbf{W_{x}}(t)$
is a standard Wiener process over the state space of $\mathbf{x}(t)$.
$\Sigma_{\mathbf{x}}$ may be a diagonal matrix of variances ($\sigma_{V}^{2}$,
$\sigma_{m_{i}}^{2}$ and $\sigma_{h_{i}}^{2}$) corresponding to
each component of the state vector. 

Typically, we assume that the above model is coupled to a measurement
``device'', which permits indirect observations of the hidden state
$\mathbf{x}(t)$: 

\begin{equation}
\mathbf{y}(t)=\mathbf{g}(\mathbf{x}(t),\zeta(t))\label{eq:obs}
\end{equation}
where $\zeta(t)$ is an observation noise vector. In the simplest
case, the vector of observations $\mathbf{y}(t)$ is one-dimensional
and it may consist of noisy measurements of the membrane potential:

\begin{equation}
y(t)=V(t)+\sigma_{y}\mathcal{N}(0,1)\label{eq:obs_simple}
\end{equation}
where $\sigma_{y}$ is the standard deviation of the observation noise
and $\mathcal{N}(0,1)$ a random number sampled from a Gaussian distribution
with zero mean and standard deviation equal to unity. More complicated
non-linear, non-Gaussian observation functions may be used when, for
example, the measurements are recordings of the intracellular calcium
concentration, simultaneous recordings of the membrane potential and
the intracellular calcium concentration or simultaneous recordings
of the membrane potential from multiple sites (e.g. soma and dendrites)
of a neuron. 

Assuming that time $t$ is partitioned in a very large number $K$
of time steps $\Delta t$, such that $t\in\{t_{0},t_{1}=t_{0}+\Delta t,t_{2}=t_{0}+2\;\Delta t,\ldots,t_{K}=K\;\Delta t\}$
and the corresponding states are $\mathbf{x}\in\{\mathbf{x}_{0},\mathbf{x}_{1},\mathbf{x}_{2},\ldots,\mathbf{x}_{K}\}$,
we can approximate the solution to Eq. \ref{eq:stoch} using the following
difference equation: 

\begin{equation}
\mathbf{x}_{k+1}=\mathbf{x}_{k}+\mathbf{f}(\mathbf{x}_{k},t_{k})\Delta t+\sqrt{\Sigma_{\mathbf{x}}}(\mathbf{W}_{\mathbf{x},k+1}-\mathbf{W}_{\mathbf{x},k})\label{eq:diff}
\end{equation}
where $\mathbf{W}_{\mathbf{x},k+1}-\mathbf{W}_{\mathbf{x},k}=\sqrt{\Delta t}\;\mathbf{\xi}_{k}$
and $\mathbf{\xi}_{k}$ is a random vector with components sampled
from a normal distribution with zero mean and unit variance. The above
expression implements a simple rule for computing the membrane potential,
activation and inactivation variables at each point $t_{k+1}$ of
the discretized time based on information at the previous time point
$t_{k}$ and it can be considered as a specific instantiation of the
Euler-Maruyama method for the numerical solution of Stochastic Differential
Equations\cite{Kloeden:1999}. 

Then, the observation model becomes:
\begin{equation}
\mathbf{y}_{k+1}=\mathbf{g}(\mathbf{x}_{k+1},\zeta_{k+1})\label{eq:obs_disc}
\end{equation}
In general, measurements do not take place at every point $t_{k}$
of the discretized time, but rather at intervals of $\Delta k$ time
steps (depending on the resolution of the measurement device), thus
generating a total of $K/\Delta k$ measurements. For simplicity in
the above description, we have assumed that $\Delta k=1$. However,
all the models we consider in the Results section assume $\Delta k>1$. 

In terms of probability density functions, the non-linear state-space
model defined by Eqs. \ref{eq:diff} and \ref{eq:obs_disc} (known
as the \textit{dynamics model}\} and the \textit{observation model},
respectively) can be written as:

\begin{eqnarray}
\mathbf{x}_{k+1} & \sim & p(\cdot|\mathbf{x}_{k})\label{eq:dyndens}\\
\mathbf{y}_{k+1} & \sim & p(\cdot|\mathbf{x}_{k+1})\label{eq:obsdens}
\end{eqnarray}
where the initial state $\mathbf{x}_{0}$ is distributed according
to a prior density $p(\mathbf{x}_{0})$. The above formulas are known
as the \textit{state transition} and \textit{observation} densities,
respectively\cite{Cappe:2007}.

\subsection*{Simulation-Based Filtering and Smoothing}

In many inference problems involving state-space models, a primary
concern is the sequential estimation of the following two conditional
probability densities\cite{Kitagawa1998}: (a) $p(\mathbf{x}_{k}|\mathbf{y}_{1:k})$
and (b) $p(\mathbf{x}_{k}|\mathbf{y}_{1:K})$, where $\mathbf{y}_{1:k}=\{\mathbf{y}_{1},...,\mathbf{y}_{k}\}$,
i.e. the set of observations (for example, a sequence of measurements
of the membrane potential) up to the time point $t_{k}$. Density
(a), known as the \textit{filter} density, models the distribution
of state $\mathbf{x}_{k}$ given all observations up to and including
the time point $t_{k}$, while density (b), known as the \textit{smoother}
density, models the distribution of state $\mathbf{x}_{k}$ given
the whole set of observations up to the final time point $t_{K}$. 

In principle, the filter density can be estimated recursively at each
time point $t_{k}$ using Bayes' rule appropriately\cite{Kitagawa1998}: 

\begin{equation}
p(\mathbf{x}_{k}|\mathbf{y}_{1:k})=\frac{p(\mathbf{y}_{k}|\mathbf{x}_{k})}{p(\mathbf{y}_{k}|\mathbf{y}_{1:k-1})}\int p(\mathbf{x}_{k}|\mathbf{x}_{k-1})p(\mathbf{x}_{k-1}|\mathbf{y}_{1:k-1})d\mathbf{x}_{k-1}\label{eq:filter}
\end{equation}
where $p(\mathbf{x}_{k}|\mathbf{x}_{k-1})$ and $p(\mathbf{y}_{k}|\mathbf{x}_{k})$
are the state transition and observation densities, respectively,
and $p(\mathbf{x}_{k-1}|\mathbf{y}_{1:k-1})$ is the filter density
at the previous time step $t_{k-1}$. 

Then, the smoother density can be obtained by using the following
general recursive formula: 

\begin{equation}
p(\mathbf{x}_{k}|\mathbf{y}_{1:K})=p(\mathbf{x}_{k}|\mathbf{y}_{1:k})\int\frac{p(\mathbf{x}_{k+1}|\mathbf{x}_{k})p(\mathbf{x}_{k+1}|\mathbf{y}_{1:K})}{p(\mathbf{x}_{k+1}|\mathbf{y}_{1:k})}d\mathbf{x}_{k+1}\label{eq:smoother}
\end{equation}
which evolves backwards in time and makes use of the pre-calculated
filter, $p(\mathbf{x}_{k}|\mathbf{y}_{1:k})$. Given either of the
above posterior densities, we can compute the expectation of any useful
function of the hidden model state as: 

\begin{equation}
\bar{h}_{k}=\int h(\mathbf{x}_{k})p(\mathbf{x}_{k}|\cdot)d\mathbf{x}_{k}\label{eq:exp}
\end{equation}
where $p(\mathbf{x}_{k}|\cdot)$ is either the filter or the smoother
density. Common examples of $h(\mathbf{x}_{k})$ are $\mathbf{x}_{k}$
itself (giving the mean $\mathbf{\bar{x}}_{k}$) and the squared difference
from the mean (giving the covariance of $\mathbf{x}_{k}$). 

In practice, the computations defined by the above formulas can be
performed analytically only for linear Gaussian models using the Kalman
smoother/filter and for finite state-space hidden Markov models. For
non-linear models, the extended Kalman filter is a popular approach,
which however can fail when non-Gaussian or multimodal density functions
are involved\cite{Cappe:2007}. A more generally applicable, albeit
computationally more intensive approach, approximates the filter and
smoother densities using Sequential Monte Carlo (SMC) methods, also
known as \textit{particle filters}\cite{Cappe:2007,Doucet:2001}.
Within the SMC framework, the filter density at each time point is
approximated by a large number $N$ of discrete samples or \textit{particles},
$\{\mathbf{x}_{k}^{(1)},\ldots,\mathbf{x}_{k}^{(N)}\}$, and associated
non-negative importance weights, $\{w_{k}^{(1)},\ldots,w_{k}^{(N)}\}$: 

\begin{equation}
p(\mathbf{x}_{k}|\mathbf{y}_{1:k})\approx\sum_{j=1}^{N}w_{k}^{(j)}\delta(\mathbf{x}_{k},\mathbf{x}_{k}^{(j)})\qquad,\qquad\sum_{j=1}^{N}w_{k}^{(j)}=1
\end{equation}
where $\delta(\mathbf{x}_{k},\mathbf{x}_{k}^{(j)})$ is the Dirac
delta function centered at the $j^{th}$ particle, $\mathbf{x}_{k}^{(j)}$. 

Given an initial set of particles sampled from a prior distribution
and their associated weights, a simple update rule involves the following
steps\cite{Kitagawa1998}: 
\begin{description}
\item [{Step~1:}] For $j=1,\ldots,N$, sample a new set of particles from
the \textit{proposal transition density function}, $q(\mathbf{x}_{k+1}^{(j)}|\mathbf{x}_{k}^{(j)},\mathbf{y}_{k+1})$.
In general, one has enormous freedom in choosing the form of this
density and even condition it on future observations, if these are
available (see, for example, \cite{Leeuwen:2010}). However, the simplest
(and a quite common) choice is to use the transition density as the
proposal, i.e. $q(\mathbf{x}_{k}|\mathbf{x}_{k-1},\mathbf{y}_{k})=p(\mathbf{x}_{k}|\mathbf{x}_{k-1})$.
This is the approach we follow in this paper.
\item [{Step~2:}] For each new particle $\mathbf{x}_{k+1}^{(j)}$, evaluate
the importance weight:
\begin{equation}
W_{k+1}^{(j)}=w_{k}^{(j)}p(\mathbf{y}_{k+1}|\mathbf{x}_{k+1}^{(j)})\frac{p(\mathbf{x}_{k+1}^{(j)}|\mathbf{x}_{k}^{(j)})}{q(\mathbf{x}_{k+1}^{(j)}|\mathbf{x}_{k}^{(j)},\mathbf{y}_{k+1})}
\end{equation}
Notice that when $q(\mathbf{x}_{k}^{(j)}|\mathbf{x}_{k-1}^{(j)},\mathbf{y}_{k})=p(\mathbf{x}_{k}^{(j)}|\mathbf{x}_{k-1}^{(j)})$,
then the computation of the importance weights is significantly simplified,
i.e. $W_{k+1}^{(j)}=w_{k}^{(j)}p(\mathbf{y}_{k+1}|\mathbf{x}_{k+1}^{(j)})$. 
\item [{Step~3:}] Normalize the computed importance weights, by dividing
each of them with their sum, i.e. 
\begin{equation}
w_{k+1}^{(j)}=\frac{W_{k+1}^{(j)}}{\sum_{j=1}^{N}W_{k+1}^{(j)}}
\end{equation}
The derived set of weighted samples $\{\mathbf{x}_{k+1}^{(j)},w_{k+1}^{(j)}\}$
is considered an approximation of the filter density $p(\mathbf{x}_{k+1}|\mathbf{y}_{k+1})$. 
\end{description}
In practice, the above algorithm is augmented with a re-sampling step
(preceding Step 1), during which $N$ particles are sampled from the
set of weighted particles computed at the previous iteration with
probabilities proportional to their weights\cite{Cappe:2007,Doucet:2001}.
All re-sampled particles are given weights equal to $1/N$. This step
results in discarding particles with small weights and multiplying
particles with large weights, thus compensating for the gradual degeneration
of the particle filter i.e. the situation where all particles but
one have weights equal to zero. For performance reasons, the resampling
step may be applied only when the effective number of particles drops
below a threshold value, e.g. $N_{thr}=N/2$. An estimation of the
effective number of particles is given by
\begin{equation}
\hat{N}_{eff}=\frac{1}{\sum_{j=1}^{N}w_{k+1}^{(j)}{}^{2}}
\end{equation}

The above filter can be extended to a fixed-lag smoother, if instead
of resampling just the particles at the current time step, we store
and resample all particles up to $L$ time steps before the current
time step, i.e. $\{\mathbf{x}_{k-L}^{(j)},\ldots,\mathbf{x}_{k-1}^{(j)},\mathbf{x}_{k}^{(j)}\}$\cite{Kitagawa1998}.
The resampled particles can be considered a realization from a posterior
density $p(\mathbf{x}_{k}|\mathbf{y}_{1:k+L})$, which is an approximation
of the smoother density $p(\mathbf{x}_{k}|\mathbf{y}_{1:K})$, for
sufficiently large values of $L$. 

Within this Monte Carlo framework, the expectation in Eq. \ref{eq:exp}
can be approximated as: 
\begin{equation}
\bar{h}_{k}\approx\sum_{j=1}^{N}w_{k}^{(j)}h(\mathbf{x}_{k}^{(j)})
\end{equation}
for a large number $N$ of weighted samples.

\subsection*{Simultaneous Estimation of Hidden States and Parameters}

It is possible to apply the above standard filtering and smoothing
techniques to parameter estimation problems involving state-space
models. The key idea\cite{Kitagawa1998} is to define an extended
state vector $\mathbf{z}_{k}$ by augmenting the state vector $\mathbf{x}_{k}$
with the model parameters, i.e. $\mathbf{z}_{k}=(\mathbf{\theta}_{k},\mathbf{x}_{k})^{T}$.
Then, the time evolution of the extended state-space model becomes:
\begin{equation}
\mathbf{z}_{k+1}=\left(\begin{array}{c}
\theta_{k+1}\\
\mathbf{x}_{k+1}
\end{array}\right)=\left(\begin{array}{c}
\theta_{k}\\
\mathbf{x}_{k}+\mathbf{f}(\mathbf{x}_{k},t_{k})\Delta t+\sqrt{\Sigma_{\mathbf{x}}\Delta t}\xi_{k}
\end{array}\right)\label{eq:extend}
\end{equation}
 while the observational model remains unaltered:
\begin{equation}
\mathbf{y}_{k+1}=\mathbf{G}(\mathbf{z}_{k+1},\zeta_{k+1})=\mathbf{g}(\mathbf{x}_{k+1},\zeta_{k+1})
\end{equation}
The marginal posterior density of the parameter vector $\theta_{k}$
is given by:
\begin{equation}
p(\theta_{k}|\mathbf{y}_{1:K})=\int p(\mathbf{z}_{k}|\mathbf{y}_{1:K})d\mathbf{x}_{k}=\int p(\theta_{k},\mathbf{x}_{k}|\mathbf{y}_{1:K})d\mathbf{x}_{k}
\end{equation}
and, subsequently, the expectation of any function of $\theta_{k}$
can be computed as in Eq. \ref{eq:exp}: 
\begin{equation}
\bar{h}_{k}=\int h(\theta_{k})p(\theta_{k}|\mathbf{y}_{1:K})d\theta_{k}
\end{equation}
Furthermore, given a set of particles and associated weights, which
approximate the smoother density $p(\mathbf{z}_{k}|\mathbf{y}_{1:K})$
as outlined in the previous section, i.e. $\{\mathbf{z}_{k}^{(j)},w_{k}^{(j)}\}=\{\mathbf{x}_{k}^{(j)},\theta_{k}^{(j)},w_{k}^{(j)}\}$
for $j=1,\ldots,N$, the above expectation can be approximated as:
\begin{equation}
\bar{h}_{k}\approx\sum_{j=1}^{N}w_{k}^{(j)}h(\theta_{k}^{(j)})
\end{equation}
for large $N$. 

Under this formulation, parameter estimation, which is traditionally
treated as an optimization problem, is reduced to an integration problem,
which can be tackled using filtering and smoothing methodologies for
state-space models, a well-studied subject in the field of Computational
Statistics.

\subsection*{Connection to Evolutionary Algorithms}

It should be emphasized that although in Eq. \ref{eq:extend} the
parameter vector $\mathbf{\theta}_{k}$ was assumed constant, i.e.
$\mathbf{\theta}_{k+1}=\mathbf{\theta}_{k}$, the same methodology
applies in the case of parameters that are naturally evolving in time,
such as a time-varying externally injected current $I_{ext}(t)$.
A particularly interesting case arises when an artificial evolution
rule is imposed on a parameter vector, which is otherwise constant
by definition. Such a rule allows sampling new parameter vectors based
on samples at the previous time step, i.e. $\theta_{k+1}\sim\mathrm{p}(\cdot|\theta_{k})$,
and generating a sequence $\{\theta_{0},\theta_{1},\ldots\}$, which
explores the parameter space and, ideally converges in a small optimal
subset of it, after a sufficiently large number of iterations. It
is at this point that the opportunity to use techniques borrowed from
the domain of Evolutionary Algorithms arises. Here, we assume that
the artificial evolution of the parameter vector $\mathbf{\theta}_{k}$
is governed by a version of the Covariance Matrix Adaptation algorithm\cite{Igel:2007},
a well-known Evolution Strategy, although the modeler is free to make
other choices (e.g. Differential Evolution\cite{Price:2005}). For
the $j^{th}$ particle, we write:
\begin{equation}
\theta_{k+1}^{(j)}=\eta_{k+1}^{(j)}+s_{k+1}^{(j)}\sqrt{Q_{k+1}}\lambda_{k+1}^{(j)}\label{eq:updtheta}
\end{equation}
where $\lambda_{k+1}^{(j)}$ is a random vector with elements sampled
from a normal distribution with zero mean and unit variance. $\eta_{k+1}^{(j)}$
and $Q_{k+1}$ are a mean vector and covariance matrix respectively,
which are computed as follows:
\begin{eqnarray}
\eta_{k+1}^{(j)} & = & (1-a)\theta_{k}^{(j)}+a\hat{E}[\theta_{k}]\label{eq:meantheta}\\
Q_{k+1} & = & (1-b)Q_{k}+b\hat{C}ov[\theta_{k}]\label{eq:covtheta}
\end{eqnarray}
In the above expressions, a and $b$ are small adaptation constants
and $\hat{E}[\cdot]$ and $\hat{C}ov[\cdot]$ are the expectation
and covariance of the weighted sample of $\theta_{k}$, respectively.
$s_{k+1}^{(j)}$ is a scale parameter that evolves according to a
log-normal update rule:
\begin{equation}
s_{k+1}^{(j)}=s_{k}^{(j)}\exp(c\phi_{k+1}^{(j)})\label{eq:lognorm}
\end{equation}
where $c$ is a small adaptation constant and $\phi_{k+1}^{(j)}\sim\mathcal{N}(0,1)$
is a normally distributed random number with zero mean and unit variance. 

According to Eq. \ref{eq:updtheta}, the parameter vector $\theta_{k+1}^{(j)}$
is sampled at each iteration of the algorithm from a multivariate
normal distribution, which is centered at $\eta_{k+1}^{(j)}$ and
has a covariance matrix equal to $s_{k+1}^{(j)}{}^{2}Q_{k+1}$:
\begin{equation}
\theta_{k+1}^{(j)}\sim\mathcal{N}(\eta_{k+1}^{(j)},s_{k+1}^{(j)}{}^{2}Q_{k+1})
\end{equation}
Both $\eta_{k+1}^{(j)}$ and $Q_{k+1}$ are slowly adapting to the
sample mean $\hat{E}[\theta_{k}]$ and covariance $\hat{C}ov[\theta_{k}]$,
with an adaptation rate determined by the constants $a$ and $b$.
Notice that by switching off the adaptation process (i.e. by setting
$a=b=c=0$), $\theta_{k+1}^{(j)}$ evolves according to a multivariate
Gaussian distribution, which is centered at the previous parameter
vector and has a covariance matrix equal to $s_{0}^{(j)}{}^{2}Q_{0}$:
\begin{equation}
\theta_{k+1}^{(j)}\sim\mathcal{N}(\theta_{k}^{(j)},s_{0}^{(j)}{}^{2}Q_{0})\label{eq:bivnorm}
\end{equation}
Therefore, given an initial set of weighted particles $\{\mathbf{z}_{0}^{(j)},w_{0}^{(j)}\}=\{s_{0}^{(j)},\theta_{0}^{(j)},\mathbf{x}_{0}^{(j)},w_{0}^{(j)}\}$
sampled from some prior density function and an initial covariance
matrix $Q_{0}$, which may be set equal to the identity matrix, the
smoothing algorithm presented earlier becomes: 
\begin{description}
\item [{Step~1a:}] Compute the expectation $\hat{E}[\theta_{k}]$ and
covariance $\hat{C}ov[\theta_{k}]$ of the weighted sample of $\theta_{k}$ 
\item [{Step~1b:}] For $j=1,\ldots,N$, compute the scale factor $s_{k+1}^{(j)}$
according to Eq. \ref{eq:lognorm}. Notice that this scale factor
is now part of the extended state $\mathbf{z}_{k+1}^{(j)}$ for each
particle 
\item [{Step~1c:}] For $j=1,\ldots,N$, compute the mean vector $\eta_{k+1}^{(j)}$,
as shown in Eq. \ref{eq:meantheta} 
\item [{Step~1d:}] Compute the covariance matrix $Q_{k+1}$, as shown
in Eq. \ref{eq:covtheta} 
\item [{Step~1e:}] For $j=1,\ldots,N$, sample $\theta_{k+1}^{(j)}$,
as shown in Eq. \ref{eq:updtheta} 
\item [{Step~1f:}] For $j=1,\ldots,N$, sample a new set of state vectors
from the proposal density $q(\mathbf{x}_{k+1}^{(j)}|\mathbf{x}_{k}^{(j)},\theta_{k+1}^{(j)},\mathbf{y}_{k+1})$,
thus completing sampling the extended vectors $\mathbf{z}_{k+1}^{(j)}$.
Notice that the proposal density $q(\cdot|\cdot)$ is conditioned
on the updated parameter vector $\theta_{k+1}^{(j)}$. 
\item [{Steps~2-3:}] Execute steps 2 and 3 as described previously 
\end{description}
Notice that in the algorithm outlined above, the order in which the
components of $\mathbf{z}_{k+1}^{(j)}$ are sampled is important.
First, we sample the scaling factor $s_{k+1}^{(j)}$. Then, we sample
the parameter vector $\theta_{k+1}^{(j)}$ given the updated $s_{k+1}^{(j)}$.
Finally, we sample the state vector $\mathbf{x}_{k+1}^{(j)}$ from
a proposal, which is conditioned on the updated parameter vector $\theta_{k+1}^{(j)}$.
When resampling occurs, the state vectors $\mathbf{x}_{k+1}^{(j)}$
with large importance weights are selected and multiplied with high
probability along with their associated parameter vectors and scaling
factors, thus resulting in a gradual self-adaptation process. This
self-adaptation mechanism is very common in the Evolution Strategies
literature.

\subsection*{Implementation}

The algorithm described in the previous section was implemented in
MATLAB and C (source code available as Supplementary Material; unmaintained
FORTRAN code is also available upon request from the first author)
and tested on parameter inference problems using simulated or actual
electrophysiological data and a number of Hodgkin-Huxley-type models:
(a) a single-compartment model (derived from the classic Hodgkin-Huxley
model of neural excitability) containing a leakage, transient sodium
and delayed rectifier potassium current, (b) a two-compartment model
of a cat spinal motoneuron\cite{Booth:1997uq} and (c) a model of
a B4 motoneuron in the Central Nervous System of the pond snail \textit{Lymnaea
stagnalis}\cite{Straub:2001kx}, which was developed as part of this
study. Each of these models is described in detail in the Results
section. Models (a) and (b) were used for generating noisy voltage
traces at a sampling rate of $10KHz$ (one sample every $0.1ms$).
The simulated data was subsequently used as input to the algorithm
in order to estimate a large number of parameters; typically, maximal
conductances of ionic currents, reversal potentials, the parameters
governing the activation and inactivation kinetics of ionic currents,
as well as the levels of intrinsic and observation noise. Estimated
parameter values were subsequently compared against the true parameter
values in the model. The MATLAB environment was used for visualization
and analysis of simulation results. For the estimation of the unknown
parameters in model (c), actual electrophysiological data were used,
as described in the next section. 

Prior information was incorporated in the smoother by assuming that
parameter values were not allowed to exceed well-defined upper or
lower limits (see Tables 1, 2 and 3). For example, maximal conductances
never received negative values, while time constants were always larger
than zero. At the beginning of each simulation, the initial population
of particles was uniformly sampled from within the acceptable range
of parameter values and, during each simulation, parameters were forced
to remain within their pre-defined limits. 

All simulations were performed on an Intel dual-core i5 processor
with 4 GB of memory running Ubuntu Linux. The number of particles
used in each simulation was typically $100\times D$, where $D$ was
the dimensionality of the extended state $\mathbf{z}$ (equal to the
number of free parameters and dynamic states in the model). The time
step $\Delta t$ in the Euler-Maruyama method was set equal to $0.01ms$.
The parameter $L$ of the fixed-lag smoother was set equal to $100$
(unless stated otherwise), which is equivalent to a time window $10ms$
wide (since data were sampled every $0.1ms$). The adaptation constants
$a$, $b$ and $c$ in Eqs. \ref{eq:meantheta}, \ref{eq:covtheta}
and \ref{eq:lognorm} were all set equal to $0.01$, unless stated
otherwise. Depending on the size of $D$, the complexity of the model
and the length of the (actual or simulated) electrophysiological recordings,
simulation times ranged from a few minutes up to more than $12$ hours.

\subsection*{Electrophysiology}

As part of this study, we developed a single-compartment Hodgkin-Huxley-type
model of a B4 neuron in the pond-snail \textit{Lymnaea stagnalis}\cite{Straub:2001kx}.
B4 neurons are part of the neural circuit that controls the rhythmic
movements of the feeding muscles via which the animal captures and
ingests its food. The \textit{Lymnaea} central nervous system was
dissected from adult animals (shell length $20-30mm$) that were bred
at the University of Leicester as described previously\cite{Straub:2007vn}.
All dissections were carried out in $HEPES$-buffered saline containing
(in $mM$) $50\; NaCl$, $1.6\; KCl$, $2\; MgCl_{2}$, $3.5\; CaCl_{2}$,
and $10\; HEPES$, $pH\;7.9$, in distilled water. All chemicals were
purchased from Sigma. The buccal ganglia containing the B4 neurons
were separated from the rest of the nervous system by cutting the
cerebral buccal connectives and the buccal-buccal connective was crushed
to eliminate electrical coupling between B4 neurons in the left and
right buccal ganglion. Prior to recording, excess saline was removed
from the dish and small crystals of protease type XIV were placed
directly on top of the buccal ganglia to soften the connective tissue
and aid the impalement of individual neurons. The protease crystals
were washed of after about $30s$ with multiple changes of $HEPES$-buffered
saline. The B4 neuron was visually identified based on its size and
position and impaled with two sharp intracellular electrodes filled
with a mixture of $3M$ potassium acetate and $10mM$ potassium chloride
(resistance $\sim20M\Omega$). During the recording, the preparation
was bathed in $HEPES$-buffered saline plus $1mM$ hexamethonium chloride
to block cholinergic synaptic inputs and suppress spontaneous fictive
feeding activity. 

The signals from the two intracellular electrodes were amplified using
a Multiclamp 900A amplifier (Molecular Devices), digitized at a sampling
frequency of $10kHz$ using a CED1401plus A/D converter (Cambridge
Electronic Devices) and recorded on a PC using Spike2 version 6 software
(Cambridge Electronic Devices). A custom set of instructions using
the Spike2 scripting language was used to generate sequences of current
pulses consisting of individual random steps ranging in amplitude
from $-4nA$ to $+4nA$ and a duration from $1$ to $256ms$. The
current signal was injected through one of the recording electrodes
whilst the second electrode was used to measure the resulting changes
in membrane potential.

\section*{Results}

\subsection*{Hidden States, Intrinsic and Observational Noise are Simultaneously
Estimated Using the Fixed-Lag Smoother}

The applicability of the fixed-lag smoother presented above was demonstrated
on a range of Hodgkin-Huxley-type models using simulated or actual
electrophysiological data. The first model we examined consisted of
a single compartment containing leakage, sodium and potassium currents,
as shown below:
\begin{equation}
dV=\frac{I_{ext}-G_{L}(V-E_{L})-G_{Na}m_{Na}^{3}h_{Na}(V-E_{Na})-G_{K}m_{K}^{4}(V-E_{K})}{C_{m}}dt+\sigma_{V}dW_{V}\label{mdl:HH}
\end{equation}
\begin{equation}
dm_{Na}=\frac{m_{\infty,Na}-m_{Na}}{\tau_{m_{Na}}}dt\qquad,\qquad dh_{Na}=\frac{h_{\infty,Na}-h_{Na}}{\tau_{h_{Na}}}dt\qquad,\qquad dm_{K}=\frac{m_{\infty,K}-m_{K}}{\tau_{m_{K}}}dt\label{mdl:HH2}
\end{equation}
where $C_{m}=1\mu F/cm^{2}$. Notice the absence of noise in the dynamics
of $m_{Na}$, $h_{Na}$ and $m_{K}$, which is valid if we assume
a very large number of channels (see Supplementary Material for the
case were noise is present in the dynamics of these variables). The
steady states and relaxation times of the activation and inactivation
gating variables were voltage-dependent, as shown below (e.g. \cite{Willms:1999ly}):
\begin{equation}
x_{\infty,i}=\left(1+\exp\frac{V_{H,x_{i}}-V}{V_{S,x_{i}}}\right)^{-1}\label{mdl:sigm}
\end{equation}
and
\begin{equation}
\tau_{x_{i}}=\tau_{min,x_{i}}+(\tau_{max,x_{i}}-\tau_{min,x_{i}})x_{\infty,i}\exp\left(\delta_{x_{i}}\frac{V_{H,x_{i}}-V}{V_{S,x_{i}}}\right)\label{mdl:tau}
\end{equation}
where $x\in\{m,h\}$ and $i\in\{Na,K\}$. The parameters $V_{H,x_{i}}$,
$V_{S,x_{i}}$, $\delta_{x_{i}}$, $\tau_{min,x_{i}}$and $\tau_{max,x_{i}}$
in Eqs. \ref{mdl:sigm} and \ref{mdl:tau} were chosen such that $x_{\infty,i}$
and $\tau_{x_{i}}$ fit closely the corresponding steady-states and
relaxation times of the classic Hodgkin-Huxley model of neural excitability
in the giant squid axon\cite{Koch:2005}. Observations consisted of
noisy measurements of the membrane potential, as shown in Eq. \ref{eq:obs_simple}.
The full set of parameter values in the above model is given in Table
1.

First, we used the fixed-lag smoother to simultaneously infer the
hidden states ($V$, $m_{Na}$, $h_{Na}$, $m_{K}$) and standard
deviations of the intrinsic ($\sigma_{V}$) and observation ($\sigma_{y}$)
noise based on simulated recordings of the membrane potential $V$.
These recordings were generated by assuming a time-dependent $I_{ext}$
in Eq. \ref{mdl:HH}, which consisted of a sequence of current steps
with amplitude randomly distributed between $-5\mu A/cm^{2}$ and
$20\mu A/cm^{2}$ and random duration up to a maximum of $20ms$.
Two simulated voltage recordings were generated corresponding to two
different levels of observation noise, $\sigma_{y}=0.5mV$ and $\sigma_{y}=50mV$,
respectively. The second value ($50mV$) was rather extreme and it
was chosen in order to illustrate the applicability of the method
even at very high levels of observation noise. Simulated data points
were sampled every $0.1ms$ ($10KHz$). The standard deviation of
the intrinsic noise was set at $\sigma_{V}=5mV$. The injected current
$I_{ext}$ and the induced voltage trace (for either value of $\sigma_{y}$)
were then used as input to the smoother, during the inference phase.
At this stage, all other parameters in the model (conductances, reversal
potentials, and ionic current kinetics) were assumed known, thus the
extended state vector took the form $\mathbf{z}=(s,\sigma_{V},\sigma_{y},V,m_{Na},h_{Na},m_{K})^{T}$,
where $s$ was a scale factor as in Eq. \ref{eq:updtheta}. New samples
for $s$ were taken from a log-normal distribution (Eq. \ref{eq:lognorm}),
while new samples for $\sigma_{V}$ and $\sigma_{y}$ were drawn from
an adaptive bivariate Gaussian distribution at each iteration of the
algorithm (Eq. \ref{eq:updtheta}). For each data set, smoothing was
repeated for two different values of the smoothing lag, i.e. $L=0$
and $L=100$. $L=0$ corresponds to filtering, while $L=100$ corresponds
to smoothing with a fixed lag equal to $10ms$. Our results from this
set of simulations are summarized in Fig. 1. 

We observed that at low levels of observation noise (Fig. 1A), the
inferred expectation of the voltage (solid blue and red lines) closely
matched the underlying (true) signal (solid black line). This was
true for both values of the fixed lag $L$ used for smoothing. However,
at high levels of observation noise (Fig. 1Bi), the true voltage was
inferred with high fidelity when a large value of the fixed lag ($L=100$)
was used (solid red line), but not when $L=0$ (solid blue line).
Furthermore, the inferred expectations of the unobserved dynamic variables
$m_{Na}$, $h_{Na}$ and $m_{K}$ (solid red lines in Fig. 1Bii) also
matched the true hidden time series (solid black lines in the same
figure) remarkably well, when $L=100$. 

We repeat that during these simulations an artificial update rule
was imposed on the two free standard deviations $\sigma_{V}$ and
$\sigma_{y}$, as shown in Eq. \ref{eq:updtheta}. The artificial
evolution of these parameters is illustrated in Fig. 1Ci, where the
inferred expectations of $s_{V}$ and $s_{y}$ are presented as functions
of time. These expectations converged immediately, fluctuating around
the true values of $s_{V}$ and $s_{y}$ (dashed lines in Fig. 1Ci).
This is also illustrated by the histograms in Fig. 1Cii, which were
constructed from the data points in Fig. 1Ci. We observed that the
peaks of these histograms were located quite closely to the true values
of $\sigma_{V}$ and $\sigma_{y}$ (dashed lines in Fig. 1Cii). 

In summary, the fixed-lag smoother was able to recover the hidden
states and standard deviations of the intrinsic and observation noise
in the model based on noisy observations of the membrane potential.
This was true even at high levels of observation noise, subject to
the condition that a sufficiently large smoothing lag $L$ was adopted
during the simulation.

\subsection*{Adaptive Sampling Reduces the Variance of Inferred Parameter Distributions
and Accelerates Convergence of the Algorithm}

Next, we treated two more parameters in the model as unknown, i.e.
the maximal conductances of the transient sodium ($G_{Na}$) and delayed
rectifier potassium ($G_{K}$) currents. The extended state vector,
thus, took the form $\mathbf{z}=(s,\sigma_{V},\sigma_{y},G_{Na},G_{K},V,m_{Na},h_{Na},m_{K})^{T}$.
As in the previous section, new samples for $s$ were drawn from a
log-normal distribution (Eq. \ref{eq:lognorm}), while $\sigma_{V}$,
$\sigma_{y}$, $G_{Na}$ and $G_{K}$ were sampled by default from
an adaptive multivariate Gaussian distribution at each iteration of
the algorithm (Eq. \ref{eq:updtheta}). 

In order to examine the effect of this adaptive sampling approach
on the variance of the inferred parameter distributions, we repeated
fixed-lag smoothing assuming each time that different aspects of this
adaptive sampling process were switched off, as illustrated in Fig.
2. First, we assumed that no adaptation was imposed on $s$ or the
``unknown'' noise parameters and maximal conductances, i.e. the
constants $a$, $b$ and $c$ in Eqs. \ref{eq:meantheta}-\ref{eq:lognorm}
were all set equal to zero. In this case, the multivariate Gaussian
distribution from which new samples of $\sigma_{V}$, $\sigma_{y}$,
$G_{Na}$ and $G_{K}$ were drawn from reduced to Eq. \ref{eq:bivnorm}.
In addition, we assumed that $s_{0}^{(j)}$ in the same equation was
equal to $1$, for all samples $j$. Under these conditions, the true
values of the free parameters were correctly estimated through application
of the fixed-lag smoother, as illustrated for the case of $G_{Na}$
and $G_{K}$ in Figs. 2Ai and 2Aii. 

Subsequently, we repeated smoothing assuming that the scale factor
$s$ evolved according to the log-normal update rule given by Eq.
\ref{eq:lognorm} with $c=0.01$, while $a$ and $b$ were again set
equal to $0$. As illustrated in Figs. 2Bi and 2Bii for parameters
$G_{Na}$ and $G_{K}$, by imposing this simple adaptation rule on
the multivariate Gaussian distribution from which the free parameters
in the model were sampled, we managed again to estimate correctly
their values, but this time the variance of the inferred parameter
distributions (the width of the histograms in Fig. 2Bii) was drastically
reduced.

By further letting the mean and covariance of the proposal Gaussian
distribution in Eq. \ref{eq:updtheta} adapt (by setting $a=b=0.01$
in Eqs. \ref{eq:meantheta} and \ref{eq:covtheta}), we achieved a
further decrease in the spread of the inferred parameter distributions
(Figs. 2C and 2D). Parameters $\sigma_{y}$ and $\sigma_{V}$ and
the hidden states $V$, $m_{Na}$, $h_{Na}$ and $m_{K}$ were also
inferred with very high fidelity in all cases (as in Fig. 1), but
the variance of the estimated posteriors for $\sigma_{y}$ and $\sigma_{V}$
followed the same pattern as the variance of $G_{Na}$ and $G_{K}$.

It is worth observing that when all three adaptation processes were
switched on (i.e. $a=b=c=0.01$), the algorithm converged to a single
point in parameter space within the first $1s$ of simulation, which
coincided with the true parameter values in the model (see Fig. 2D
for the case of $G_{Na}$ and $G_{K}$). At this point, the covariance
matrix $\bar{s}_{k}^{2}Q_{k}$ became very small (i.e. all its elements
were less than $10^{-8}$, although the matrix itself remained non-singular)
and the mean $\bar{\eta}_{k}$ was very close to the true parameter
vector $\theta$. We note that $\bar{s}_{k}=\hat{E}[s_{k}]$ and $\bar{\eta}_{k}=\hat{E}[\eta_{k}]$,
where $\hat{E}[\cdot]$ stands for the expectation computed over the
population of particles. In this case, it is not strictly correct
to claim that the chains in Fig. 2Di approximate the posteriors of
the unknown parameters $G_{Na}$ and $G_{K}$; since repeating the
simulation many times would result in convergence at slightly different
points clustered tightly around the true parameter values, it would
be more reasonable to claim that these optimal points are random samples
from the posterior parameter distribution and they can be treated
as estimates of its mode. 

Depending on the situation, one may wish to estimate the full posteriors
of the unknown parameters or just an optimal set of parameter values,
which can be used in a subsequent predictive simulation. In Fig 3A,
we examined in more detail how the scale factor $s_{k}$ affects the
variance of the final estimates, assuming that $a=b=c=0.01$. We repeat
that each particle $j$ contains $s_{k}$ as a component of its extended
state. Each scaling factor $s_{k}^{(j)}$ is updated at each iteration
of the algorithm following a lognormal rule (Eq. \ref{eq:lognorm},
Step 1b of the algorithm in the Methods section). Sampling new parameter
vectors is conditioned on these updated scaling factors (Eq. \ref{eq:updtheta},
Step 1e of the algorithm). When at a later stage weighting (and resampling)
of the particles occurs, the scaling factors that are associated with
high-weight parameters and hidden states are likely to survive into
subsequent iterations (or ``generations'') of the algorithm. During
the course of this adaptive process, the scaling factors $s_{k}^{(j)}$
are allowed to fluctuate only within predefines limits, similarly
to the other components of the extended state vector. 

In Fig. 3Ai, we demonstrate the case where the scaling factors $s_{k}^{(j)}$
were allowed to take values from the prior interval $[0,2]$. We observed
that during the course of the simulation, the average value of the
scaling factor, $\bar{s}_{k}$, decreased gradually towards $0$ and
this was accompanied by a dramatic decrease in the variance of the
inferred parameters $G_{Na}$ and $G_{K}$, which eventually ``collapsed''
to a point in parameter space located very close to their true values.
This situation was the same as the one illustrated in Fig. 2D. Notice
that although $\bar{s}_{k}$ decreased towards zero, it never actually
took this value; it merely became very small ($\sim0.01$). When we
used a prior interval for $s_{k}^{(j)}$ with non-zero lower bound
(i.e. $[0.15,2${]}; see Fig. 3Aii), the final estimates had a larger
variance, providing an approximation of the full posteriors of the
``unknown'' parameters $G_{Na}$ and $G_{K}$. Thus, controlling
the lower bound of the prior interval for the scaling factors $s_{k}^{(j)}$
provides a simple method for controlling the variance of the final
estimates. Notice that the variance of the final estimates also depends
on the number of particles (Fig. 3B). A smaller number of particles
resulted in a larger variance of the estimates (compare Fig. 3Bi to
Fig. 3Bii). However, when a large number of particles was already
in use, further increasing their number did not significantly affect
the variance of the estimates or rte of convergence (compare Fig.
3Bii to Fig. 3Aii), indicating the presence of a ceiling effect.

The adaptive sampling of the scaling factors $s_{k}^{(j)}$ further
depends on parameter $c$ in Eq. \ref{eq:lognorm}, which determines
the width of the lognormal distribution from which new samples are
drawn. The value of this parameter provides a simple way to control
the rate of convergence of the algorithm; larger values of $c$ resulted
in faster convergence (compare Fig. 4A to Fig 4B). The rate of convergence
also depends on the number of particles in use (compare Fig. 4A to
Fig. 4C), although it is more sensitive to changes in parameter $c$;
dividing the value of $c$ by $2$ (Fig. 4B) had a larger effect on
the rate of convergence than dividing the number of particles by $10$
(Fig. 4C). 

In summary, by assuming an adaptive sampling process for the unknown
parameters in the model, we managed to achieve a significant reduction
in the spread of the inferred posterior distributions of these parameters.
Furthermore, adjusting the prior interval and adaptation rate $c$
of the scaling factors $s_{k}^{(j)}$ provides a straightforward way
to control the variance of the estimated posteriors and the rate of
convergence of the algorithm. Alternatively, we could have set $s_{k}^{(j)}=constant$,
i.e. set it to the same constant value for all particles $j$ and
time steps $k$ (as in Fig. 2A). However, by permitting $s_{k}^{(j)}$
to adapt within a predefined interval, we potentially allow this parameter
and, thus, the covariance matrices $s_{k}^{(j)}{}^{2}Q_{k}$ take
large values, which in turn would permit the algorithm to escape local
optima in the parameter space. For example, the time profiles of $\bar{s}_{k}$
in Figs. 3 and 4 indicate that, early during the simulations, $ $this
quantity had relatively large values, which were associated with large
variances of the posterior parameter estimates. During this initial
period, the algorithm has the potential to ``jump'' away from local
optima and towards more optimal regions of the parameter space. One
may see, here, a distant analogy to simulated annealing, where a fictitious
``temperature'' control variable is gradually decreased, thus allowing
the system to escape local minima and gradually settle to more optimal
regions of the energy landscape.

\subsection*{Increasing Observation Noise Reduces the Accuracy and Precision of
the Fixed-Lag Smoother}

In a subsequent stage, we treated as unknown two more parameters in
the model, i.e. the reversal potentials for the sodium and potassium
currents, $E_{Na}$ and $E_{K}$, respectively. Thus, the extended
state vector became $\mathbf{z}=(s,\sigma_{V},\sigma_{y},G_{Na},G_{K},E_{Na},E_{K},V,m_{Na},h_{Na},m_{K})^{T}$.
This time, we wanted to examine how increasing levels of observation
noise (i.e. the value of parameter $\sigma_{y}$) affect the inference
of unknown quantities in the model based on the fixed-lag smoother.
For this reason, we repeated smoothing on four simulated data sets
(i.e. recordings of membrane potential and the associated $I_{ext}$)
corresponding to increasing values of the standard deviation of the
observation noise $\sigma_{y}$, i.e. $0.5mV$, $5mV$, $25mV$ and
$50mV$.

The results from this set of simulations are summarized in Fig. 5.
For $\sigma_{y}=0.5mV$, the expectations of the four parameters $G_{Na}$,
$G_{K}$, $E_{Na}$ and $E_{K}$ (red solid lines in Figs. 5Ai-iv)
eventually converged to their true values (dashed lines in the aforementioned
figures). For $\sigma_{y}=50mV$, the expectations of these parameters
(light red solid lines in Figs. 5Ai-iv) also converged, although the
expectations for $G_{Na}$ (Fig. 5Ai) and, to a lesser degree, $G_{K}$
(Fig. 5Aii) deviated noticeably from their true values. As expected,
at higher levels of noise, the variance of the final estimates was
larger, although the rate of convergence did not seem to be affected,
due to the large number of particles we used ($N=1100$; see ceiling
effect in Fig. 3Bii). The inferred parameters $\sigma_{V}$ and $\sigma_{y}$
(not illustrated for clarity) followed a similar convergence pattern.

In Fig. 5B, we show, for each tested value of $\sigma_{y}$, the box
plots of the above four parameters, which were computed from the data
points (as in Fig. 5A) corresponding to time $t\ge1s$. For each parameter
and each value of $\sigma_{y}$, the data were first normalized as
follows:
\begin{equation}
\tilde{x}_{k}=\frac{\bar{x}{}_{k}-x_{true}}{\sum_{k=1}^{K}\bar{x}_{k}}\label{eq:normrule}
\end{equation}
where $x\in\{G_{Na},G_{K},E_{Na},E_{K}\}$. The box plots in Fig.
5B were constructed from the normalized data points $\tilde{x}_{k}$.
The above normalization was necessary since it made possible the comparison
between different data sets, each characterized by its own mean, variance
and unit of measurement. In the box plots in Fig. 5B, zero (i.e. the
dashed lines) corresponds to the true parameter values, while discrepancies
from the true parameter values along the y-axis are given in relation
to the average $\sum_{k=1}^{K}\bar{x}_{k}$. We may observe that for
very low levels of observation noise ($\sigma_{y}=0.5mV$), the posteriors
of the four examined parameters were clustered tightly around their
true values, but for larger levels of noise ($\sigma_{y}=5$, $25$
and $50mV$), we observed larger discrepancies from the true parameter
values and broader inferred posteriors. The parameters following more
noticeably this trend were the conductances $G_{Na}$ and $G_{K}$,
while $E_{Na}$ and, particularly, $E_{K}$ were less affected. This
indicates that smoothing is more sensitive to changes in some model
parameters than others and this is why these parameters were tightly
controlled. In summary, increasing the levels of measurement noise
(i.e. the value of parameter $\sigma_{y}$) decreased the accuracy
and precision of the algorithm, but it did not significantly affect
the rate of convergence due to the large number of particles used
during the simulations.

\subsection*{High-Dimensional Inference Problems are Resolved Given Sufficiently
Informative Priors}

At the next stage, we treated all parameters in the model (a total
of 23 parameters; see Table 1) as unknown. Therefore, the extended
state vector took the following ($28$-dimensional) form:
\[
\mathbf{z}=(s,\sigma_{V},\sigma_{y},G_{L},G_{i},E_{L},E_{i},V_{H,x_{i}},V_{S,x_{i}},\tau_{min,x_{i}},\tau_{max,x_{i}},\delta_{x_{i}},V,m_{Na},h_{Na},m_{K})^{T}
\]
where $i\in\{Na,K\}$ and $x\in\{m,h\}$. These parameters included
the standard deviations of intrinsic and observation noise ($\sigma_{V}$
and $\sigma_{y}$, respectively), the maximal conductances $G_{i}$
and reversal potentials $E_{i}$ of all currents in the model and
the parameters controlling the steady-states and relaxation times
of activation and inactivation for the sodium and potassium currents
($V_{H,x_{i}}$, $V_{S,x_{i}}$, $\tau_{min,x_{i}}$, $\tau_{max,x_{i}}$
and $\delta_{x_{i}}$). The results from this simulation are illustrated
in Fig. 6.

We observed that the true signal (membrane potential) was inferred
with very high fidelity (Fig. 6Ai). The sodium activation $m_{Na}$
was also recovered with very high accuracy, while estimation of the
hidden states $h_{Na}$ and $m_{K}$ (sodium inactivation and potassium
activation, respectively) was also satisfactory (despite significant
deviations, the general form of the true hidden states was recovered
without any observable impact on the dynamics of the membrane potential),
as shown in Fig. 6Aii. Among the $23$ estimated parameters, we illustrate
(in Figs. 6B and 6C) the estimated posteriors for the reversal potential
of sodium $E_{Na}$ (Fig. 6B) and for parameters $\tau_{max,m_{Na}}$
(Figs. 6Ci,ii) and $\tau_{max,m_{K}}$ (Figs. 6Ciii,iv), which control
the activation of sodium and potassium currents, respectively. We
focus on these parameters, because they represent three different
characteristic cases. The posteriors of parameters $E_{Na}$ and $\tau_{max,m_{Na}}$
are unimodal (see Figs. 6Bii and 6Cii) and they were estimated with
relatively high accuracy. Particularly, the posterior for $\tau_{max,m_{Na}}$
was estimated with very high precision and accuracy, despite its broad
prior interval (the y-axis in Fig. 6Ci and the x-axis in Fig. 6Cii).
On the other hand, the estimated posterior of $\tau_{max,m_{K}}$
covered a large part of its prior interval (the y-axis in Fig. 6Ciii
and the x-axis in Fig. 6Civ), its main mode was located at a slightly
larger value than the true parameter value, while at least two minor
modes seem to be present near the upper bound of the prior interval
(the arrow in Fig. 6Civ). These results reiterate our previous conclusion
that smoothing may be particularly sensitive to some parameters, but
not to others. The posteriors of parameters in the former category
are very precise and narrow (as in the case of $E_{Na}$ and, especially,
$\tau_{max,m_{Na}}$), while the parameters in the latter category
are characterized by broader posteriors. Also, we can observe that
the fixed-lag smoother has the capability to provide a global approximation
of the unknown posteriors, including their variance and the location
of major and minor modes (i.e. global and local optima). An overview
of all inferred posteriors is given by the box plot in Fig. 6D, which
was constructed after all data (as in Figs. 6Bi, 6Ci and 6Ciii) were
normalized according to Eq. \ref{eq:normrule}. Again, it may be observed
that while some of the estimated parameter posteriors are quite precise
and accurate, such as $\sigma_{y}$ (parameter $\#2$), $E_{K}$ (parameter
$\#8$) and $V_{H,m_{Na}}$ (parameter $\#9$), others are less precise
and accurate, such as the maximal conductances (parameters $\#3$
to $\#5$), $\tau_{max,h_{Na}}$ (parameter $\#19$) and $\delta_{h_{Na}}$
(parameter $\#22$).

The simulation results presented above were obtained by assuming a
prior interval for the scaling factors $s_{k}^{(j)}$ equal to $[0.15,10]$.
When we repeated the simulation using the prior interval $[0,10]$,
the true underlying membrane potential was again inferred with very
high fidelity (Fig. 7Ai), while the hidden states $m_{Na}$, $h_{Na}$
and $m_{K}$ were also estimated with sufficient accuracy (Fig. 7Aii).
In this case, however, the estimates of the ``unknown'' parameters
converged to single points in parameter space (as illustrated, for
example, for parameters $E_{Na}$, $\tau_{max,m_{Na}}$ and $\tau_{max,m_{K}}$
in Figs. 7Bi-ii), which fall within the support of the posteriors
illustrated in Figs. 6B and 6C. The activation and inactivation steady
states (Fig. 7Ci, red solid lines) and relaxation times (Fig. 7Cii,
red solid lines) as functions of voltage, which were computed from
these estimates, were also similar to their corresponding true functions,
with the curves for $\bar{\tau}_{h_{Na}}$ and $\bar{\tau}_{m_{K}}$
manifesting the largest deviation from truth (black solid lines in
Figs. 7Ci,ii). An overview of the estimated parameter values (after
normalizing using Eq. \ref{eq:normrule}) is given in Fig. 7Di. As
stated previously, some estimates were close to their true counterparts,
while others were not. For example, the activation of the sodium current
$m_{Na}$ (Fig. 7Aii) and its steady state $m_{\infty,Na}$ (Fig.
7Ci), which are important for the correct onset of the action potentials,
were inferred with relatively high accuracy. On the other hand, larger
errors were observed, for example, in the inference of sodium inactivation
($h_{Na}$; Fig. 7Aii) or in the estimation of $G_{Na}$ (parameter
$\#4$; Fig. 7Di), the maximal conductance for the sodium current. 

Given the fact that the data on which inference was based (a single
noisy recording of the membrane potential) was of much lower dimensionality
than the extended state we aimed to infer, the observed discrepancies
between inferred and true model quantities were unlikely to vanish
unless we imposed more strict constraints on the model. When we repeated
the previous simulation using more narrow prior intervals for some
of the parameters controlling the kinetics of the sodium and potassium
currents in the model (see red dashed boxes in Fig. 7Dii and bold
intervals in Table 1), the estimated parameters settled closer to
their true values (Fig. 7Dii). This was true even for parameters on
which more narrow intervals were not directly applied, such as the
maximal conductances (i.e. parameters $\#3$ to $\#5$ in Fig. 7Dii),
and even when data with higher levels of observation noise wee used
(Fig. 7Dii, data points indicated with crosses; see also Fig. S3).
It is important to mention that using more narrow prior constraints
only affected the accuracy of the final estimates, not the quality
of fitting the experimental data, which in all cases was of very high
fidelity. Alternatively, we could have constrained the model by increasing
the dimensionality of the observed signal, e.g. by using simultaneously
more that one unique voltage traces (each generated under different
conditions of injected current) during smoothing. We examine the use
of multiple data sets simultaneously as input to the fixed-lag smoother
later in the Results section. 

In summary, the smoothing algorithm can be used to resolve high-dimensional
inference problems. In combination with sufficient prior information
(in the form of bounded regions within which parameters are allowed
to fluctuate; see Table 1), the fixed-lag smoother can provide estimates
of the intrinsic and observation noise, maximal conductances, reversal
potentials and kinetics of ionic currents in a single-compartment
Hodgkin-Huxley-type neuron model, based on low-dimensional noisy experimental
data.

\subsection*{Parameter Estimation in Compartmental Models is Straightforward Using
the Fixed-Lag Smoother}

Next, we tested whether the fixed-lag smoother could be successfully
applied on inference problems involving more complex models than the
one we used in the previous sections. For this reason, we focused
on a two-compartment model of a vertebrate motoneuron containing sodium,
potassium and calcium currents and intracellular calcium dynamics,
which were differentially distributed among a soma and a dendritic
compartment\cite{Booth:1997uq}. The model (modified appropriately
to include intrinsic noise terms) is summarized below: 
\begin{eqnarray}
dV_{S} & = & \frac{I_{ext,S}-G_{L}(V_{S}-E_{L})-\frac{G_{C}}{p}(V_{S}-V_{D})-I_{Na}-I_{K}-I_{K(Ca),S}-I_{CaN,S}}{C_{m}}dt+\sigma_{V_{S}}dW_{V_{S}}\label{mdl:motosoma}\\
dV_{D} & = & \frac{I_{ext,D}-G_{L}(V_{D}-E_{L})-\frac{G_{C}}{1-p}(V_{D}-V_{S})-I_{K(Ca),D}-I_{CaN,D}-I_{CaL}}{C_{m}}dt+\sigma_{V_{D}}dW_{V_{D}}\label{mdl:motodend}
\end{eqnarray}
where $V_{S}$ and $V_{D}$ is the membrane potential at the soma
and dendritic compartments, respectively, and $C_{m}=1\mu F/cm^{2}$.
The leakage conductance and reversal potential were $G_{L}=0.51mS/cm^{2}$
and $E_{L}=-60mV$, respectively. The coupling conductance was $G_{C}=0.1mS/cm^{2}$
and the ratio of the soma area to the total surface area of the cell
was $p=0.1$. The various ionic currents in the above model were as
follows: (a) a transient sodium current, $I_{Na}=G_{Na}m_{\infty,Na}^{3}h_{Na}(V_{S}-E_{Na})$,
(b) a delayed rectifier potassium current, $I_{K}=G_{K}m_{K}^{4}(V_{S}-E_{K})$,
(c) a calcium-activated potassium current, $I_{K(Ca),X}=G_{K(Ca),X}\frac{[Ca^{2+}]_{X}}{[Ca^{2+}]_{X}+K_{d}}(V_{X}-E_{K})$,
where $X\in\{S,D\}$ and $K_{d}=0.2\mu M$ (the half-saturation constant),
(d) an N-type calcium current, $I_{CaN,X}=G_{CaN,X}m_{CaN,X}^{2}h_{CaN,X}(V_{X}-E_{Ca})$,
where $X\in\{S,D\}$ and (e) an L-type calcium current, $I_{CaL}=G_{CaL}m_{CaL}(V_{D}-E_{Ca})$.
The various activation and inactivation dynamic variables in the above
model were modeled using first-order relaxation kinetics (as in Eq.
\ref{mdl:HH2}), where the various steady states were assumed to be
sigmoid functions of voltage (Eq. \ref{mdl:sigm}). Notice, that the
activation of $I_{Na}$ was assumed instantaneous and therefore, it
was given at any time by the voltage-dependent steady state $m_{\infty,Na}$.
The relaxation times for sodium inactivation and potassium activation
were also functions of voltage as in Eq. \ref{mdl:tau}:
\begin{equation}
\tau_{h_{Na}}=\tau_{max,h_{Na}}h_{\infty,Na}\exp\left(\delta_{h_{Na}}\frac{V_{H,h_{Na}}-V}{V_{S,h_{Na}}}\right)
\end{equation}
\begin{equation}
\tau_{m_{K}}=\tau_{min,m_{K}}+(\tau_{max,m_{K}}-\tau_{min,m_{K}})m_{\infty,K}\exp\left(\delta_{m_{K}}\frac{V_{H,m_{K}}-V}{V_{S,m_{K}}}\right)
\end{equation}
where the parameters $\tau_{min,x_{i}}$, $\tau_{max,x_{i}}$ and
$\delta_{x_{i}}$ (with $x\in\{m,h\}$ and $i\in\{Na,K\}$) were chosen
by fitting the above expressions to the original model in \cite{Booth:1997uq}.
The relaxation times for the remaining activation and inactivation
variables were constant. All parameters values in the model are given
in Table 2. 

The intracellular calcium concentration at either the soma or the
dendritic compartment was also modeled by a first-order differential
equation, as follows:
\begin{equation}
\frac{d[Ca^{2+}]_{X}}{dt}=f(aI_{Ca,X}-k[Ca^{2+}]_{X})\qquad,\qquad X\in\{S,D\}
\end{equation}
where $f=0.01$, $a=0.009mol(C\mu m)^{-1}$ and $k=2ms^{-1}$. The
total calcium current is $I_{Ca,S}=I_{CaN}$ at the soma ($X=S$)
and $I_{Ca,D}=I_{CaN}+I_{CaL}$ at the dendritic compartment ($X=D$).

The observation model assumed simultaneous noisy recordings of the
membrane potential from both the soma and dendritic compartments,
as follows: 
\begin{equation}
\left(\begin{array}{c}
y_{S}\\
y_{D}
\end{array}\right)=\left(\begin{array}{c}
V_{S}\\
V_{D}
\end{array}\right)+\left(\begin{array}{cc}
\sigma_{y} & 0\\
0 & \sigma_{y}
\end{array}\right)\left(\begin{array}{c}
\zeta_{S}\\
\zeta_{D}
\end{array}\right)
\end{equation}
where $\zeta_{X}\sim\mathcal{N}(0,1)$ with $X\in\{S,D\}$. Notice
that $\sigma_{y}$ is the same for both compartments. 

In the above model, the externally injected currents $I_{ext,S}$
and $I_{ext,D}$ were sequences of random current steps with duration
up to $50ms$ (instead of $20ms$ as in the single-compartment model,
due to the presence of slower currents in the two-compartment model)
and magnitude between $-5\mu A/cm^{2}$ and $20\mu A/cm^{2}$. Current
was injected in both the dendritic compartment and the soma (instead
of just in the soma), because preliminary simulations indicated that
this experimental setting facilitated parameter estimation, presumably
due to the generation of a more variable (and, thus, information-rich)
data set%
\footnote{It should be mentioned that the two-compartment model allows for the
physical separation of currents and as such it is a slightly better
approximation of a real neuron with differential expression of individual
currents in different cellular compartments. However, in no way does
it capture the full morphological complexity of a real neuron. As
such, current injection into the dendritic compartment can not be
replicated accurately in a real neuron as current injection in the
model will have a uniform effect on all currents in that compartment,
whilst current injection into the dendrite of a neuron would have
far more complex effects on dendritic currents, which potentially
would be dependent on the distance from the injection site. Thus,
whilst it would be possible, albeit challenging, to carry out dual
recordings from the soma and dendrites in a real neuron this would
not be the same as the dual current injection in the model. In this
case, application of the fixed-lag smoother on a more spatially detailed
model would be necessary (and feasible). In principle, the method
can also assimilate other types of spatial data, such as calcium imaging
data, in case recordings from multiple neuron locations are not available
(although we do not examine this case in detail in this paper).%
}. The injected currents and the induced noisy voltage traces $y_{S}$
and $y_{D}$ comprised the simulated data on which parameter estimation
was based. 

First, we aimed to infer the noise parameters and maximal conductances
of all voltage- and calcium-gated currents in the model, assuming
that the kinetics of these currents were known. This implied an extended-state
vector with $22$ components as shown below
\[
\mathbf{z}=(s,\sigma_{X},\sigma_{y},G_{Na},G_{K},G_{K(Ca),X},G_{CaN,X},G_{CaL},V_{X},[Ca^{2+}]_{X},h_{Na},m_{K},m_{CaN,X},h_{CaN,X},m_{CaL})^{T}
\]
where $X\in\{S,D\}$. The results from this simulation are illustrated
in Figs. 8 and 9. The fixed-lag smoother managed to recover the hidden
dynamic states (including the time-evolution of the intracellular
calcium; Fig. 8), the standard deviations of the intrinsic and observation
noise (Figs. 9Ai,ii) and the true values of all the gated maximal
conductances (Figs. 9Bi-iv) in the model using approximately $2s$
of simulated data and $2200$ particles. Notice that, in Figs. 8Ci-iv,
the inferred hidden gating states (dashed red lines) coincide extremely
well with the true ones (solid black lines), which is not surprising,
since the voltage-dependent kinetics of these states were assumed
known and the true membrane potential at the soma and dendritic compartment
was recovered with very high fidelity (Figs. 8Ai,ii). Also, notice
that, in Figs. 9Aii, 9Biii and 9Biv, the estimation of the standard
deviation of the intrinsic noise, $\sigma_{V_{D}}$, and the maximal
conductances of calcium and calcium-dependent currents in the dendritic
compartment ($G_{K(Ca),D}$, $G_{CaN,D}$ and $G_{CaL}$) was improved
after injecting current in both the soma and the dendritic compartment
(compare the grey solid lines, which correspond to injection in the
soma only, to the color ones in the aforementioned figures). 

In a second stage, we assumed that the kinetics of all voltage-gated
ionic currents were also unknown, implying an extended state vector
with $41$ components, as follows: 
\[
\mathbf{z}=(\ldots,G_{CaL},V_{H,x_{i}},V_{S,x_{i}},\tau_{min,x_{i}},\tau_{max,x_{i}},\delta_{x_{i}},\tau_{o,x_{i}},V_{X},\ldots)^{T}
\]
where $X\in\{S,D\}$, $x\in\{m,h\}$ and $i\in\{Na,K,CaN,CaL\}$.
Our results from this simulation are summarized in Figs. 10 and 11.
Again, the membrane potential at the soma and the dendrite were inferred
with very high fidelity (Fig. 10Ai,ii). However, the estimated hidden
dynamics of most ionic currents and intracellular calcium concentrations
in the model deviated significantly from their true counterparts (Fig.
10B,C). The expectations of all estimated parameters are illustrated
in Fig. 11Ai. As in the case of the single-compartment model, by imposing
tighter prior constraints on some of the parameters controlling the
kinetics of ionic currents in the model (see red dashed box in Fig.
11Aii and Table 2), we managed to reduce the discrepancies of the
estimates from their true values (Fig. 11Aii and Supplementary Fig.
S4). This was true even for parameters on which stricter priors were
not directly applied. The inference was completed after processing
almost $3s$ of data, as shown in Fig. 11B for the maximal conductances
of sodium and potassium currents at the soma. Interestingly, the algorithm
seems to temporarily settle at local optima (see arrows in Fig. 11B)
before ``jumping'' away and, eventually, converge at the final estimates.
The inferred voltage-dependent steady-states of the sodium, potassium
and calcium currents (Figs. 11Ci,ii) and the relaxation times for
the sodium inactivation and potassium activation (Fig. 11Ciii) were
also very similar to their true corresponding functions. The algorithm
remained operational when more noisy data were used, as illustrated
in Fig. 11Aii and in Supplementary Fig. S5. 

An interesting fact regarding the simulation results presented in
Figs. 10 and 11Ai was that, in order to obtain high-fidelity estimates
of the true membrane potential at the soma and dendritic compartment
(as shown in Figs. 10Ai,ii) we had to use more than $4100$ particles,
the number calculated by the $N=100\times\text{size of the extended state}$
rule (see Methods). In particular, we used $8200$ particles, although
we cannot exclude that a smaller number may have sufficed. After applying
more narrow prior constraints (Figs. 11Aii, B, C, S4 and S5), using
the number of particles calculated by the above simple heuristic ($4100$
in this case) was again sufficient for accurately inferring the true
membrane potential (see Fig. S4Ai,ii and S5Ai,ii). This implies that
as the complexity (and dimensionality) of the estimation problem increases,
a non-linearly growing number of particles may be required in order
to obtain acceptable results, but this situation may be compensated
for by providing highly informative priors. 

Given the large number of unknown parameters and hidden states in
combination with the low dimensionality of the data (notice that the
intracellular calcium concentration was assumed unobserved), it was
truly remarkable that the algorithm managed to recover much of the
extended state vector with relatively satisfactory accuracy. However,
it should be noted that in our simulations we assumed knowledge of
important information, such as the passive conductances $G_{L}$ and
$G_{C}$ and the reversal potentials of sodium, potassium and calcium
currents. This and the fact that the availability of prior information
in the form of more narrow parameter boundaries improved significantly
the accuracy of the final estimates emphasizes our previous conclusion
that prior information is important for the successful inference of
unknown model parameters and hidden model states using the fixed-lag
smoother. Given such information, inference in complex compartmental
models based on simultaneous recordings from several neuron locations
and, possibly, measurements of intracellular calcium, can be naturally
achieved via appropriate formulation of the extended state vector
and application of the fixed-lag smoother.

\subsection*{Parameters in a Model of an Invertebrate Motoneuron were Inferred
from Actual Electrophysiological Data Using the Fixed-Lag Smoother}

In a final set of simulations, we applied the smoother on actual electrophysiological
data in order to estimate the unknown parameters in a single-compartment
model of the B4 motoneuron from the nervous system of the pond snail,
\textit{Lymnaea stagnalis}\cite{Straub:2001kx}. This neuron is part
of a population of motoneurons, which receive rhythmic electrical
input from upstream Central Pattern Generator interneurons and in
turn innervate and control the movements of the feeding muscles via
which the animal captures and ingests its food. Previous studies in
these neurons have demonstrated the presence of a transient inward
sodium current $I_{Na}$, a delayed outward potassium current $I_{K}$
and a transient outward potassium current $I_{A}$\cite{Vehovszky:2005ys}.
A hyperpolarization-activated current $I_{h}$ was conditional on
the presence of serotonin in the solution \cite{Straub:2001kx} and,
therefore, this current was not included in this instance of the B4
model. Thus, the current conservation equation for a single-compartment
model of the B4 motoneuron (appropriately modified to include an intrinsic
noise term) took the following form:
\begin{equation}
dV=\frac{I_{ext}-G_{L}(V-E_{L})-I_{Na}-I_{K}-I_{A}}{C_{m}}dt+\sigma_{V}dW_{V}
\end{equation}
where the leakage conductance, leakage reversal potential and membrane
capacitance in the above model were estimated \textit{a priori} based
on neuron responses to negative current pulses ($G_{L}=0.11\mu S$,
$E_{L}=-65mV$ and $C_{m}=2.89nF$, respectively). The voltage-activated
currents that appear in the above expression were modeled as follows:
(a) $I_{Na}=G_{Na}m_{\infty,Na}^{3}h_{Na}(V-E_{Na})$, (b) $I_{K}=G_{K}m_{K}^{4}(V-E_{K})$
and (c) $I_{A}=G_{A}m_{A}^{4}h_{A}(V-E_{K})$, where $E_{Na}=35mV$
and $E_{K}=-67mV$ as in \cite{Vehovszky:2005ys}. The dynamic activation
and inactivation variables of these currents ($h_{Na}$, $m_{K}$,
$m_{A}$ and $h_{A}$) obeyed first-order relaxation kinetics (as
in Eq. \ref{mdl:HH2}) with voltage-dependent steady-states (Eq. \ref{mdl:sigm})
and relaxation times (Eq. \ref{mdl:tau} with $\tau_{min,x_{i}}=0$
and $\delta_{x_{i}}=0.5$), similarly to previously published neuron
models in the central nervous system of \textit{Lymnaea}\cite{Vavoulis:2010zr}.
The observation model was as in Eq. \ref{eq:obs_simple}.

The raw data we used for inferring the parameters in the above model
took the form of four independent $3.5s$-long recordings of the membrane
potential from the same B4 motoneuron. Each recording was taken while
injecting an external current in the neuron consisting of a sequence
of random steps ranging in amplitude between $-4nA$ and $+4nA$ and
with duration between $1$ and $256ms$. A particular characteristic
of the data generated under these conditions was the presence of brief
bursts of spikes, which were interrupted by relatively long intervals
of non-activity (corresponding to sub-threshold excitatory and inhibitory
current injections, respectively; see Figs. 12Ai-iv). These long intervals
of inactivity were not informative and they negatively affected the
performance of the smoother by permitting the random drift of particles
towards non-optimal regions of the parameter space (see Supplementary
Fig. S6). However, when the four recordings are considered together,
the intervals of inactivity at any single voltage trace overlap with
intervals of activity at the remaining three voltage traces, resulting
in a four-dimensional data set, where the overall intervals of inactivity
were minimized. This four-dimensional data set was used as input to
the smoother during the inference phase. 

Thus, the $42$-dimensional extended state vector became:
\[
\mathbf{\mathbf{z}=}(s,G_{i},V_{H,x_{i}},V_{S,x_{i}},\tau_{max,x_{i}},V_{k},m_{Na,k},h_{Na,k},m_{K,k},m_{A,k},h_{A,k})^{T}
\]
where $x\in\{m,h\}$, $i\in\{Na,K,A\}$ and $k\in\{1,2,3,4\}$. Notice
the presence of four groups of hidden dynamic states, \{$V_{k}$,
$m_{Na,k}$, $h_{Na,k}$, $m_{K,k}$, $m_{A,k}$, $h_{A,k}$\}, where
each group corresponds to a different voltage trace (and associated
externally injected current, $I_{ext,k}$). The evolution of all four
groups of dynamic variables was governed by a common (shared) set
of parameters. In total, we had to estimate $17$ unknown parameters.
The boundaries within which the values of these parameters were allowed
to fluctuate are given in Table 3 (indicated in bold) and they were
chosen from within the support of the posteriors in Supplementary
Fig. S7 (after a few trial-and-error simulations), which were obtained
by using the broader prior intervals given in Table 3. Notice that
the marginal distributions illustrated in Fig. S7 have large variance
and multiple modes and, although they provide a global view of the
structure of the parameter space, they cannot be used to identify
a single combination of optimal parameters values, since they do not
include any information regarding correlations between parameters.
Using the major modes of the inferred posteriors did not lead to an
accurate (or even spiking) predictive model. Thus, the estimation
was based on using more narrow prior intervals, which helped us estimate
unimodal posteriors with small variance (see Fig. 12C) and, thus,
identify a single combination of optimal parameters that could be
used in predictive simulations. We cannot prove that other optimal
combinations of parameters do not exist, but we were not able to find
any (i.e. by choosing different narrow prior intervals) after a reasonable
amount of time. Also, notice that the standard deviations of the intrinsic
and observation noise were not subject to estimation, but instead
they were given (through trial and error) the minimal fixed values
$\sigma_{V}=0.3mV$ and $\sigma_{y}=1mV$, respectively. If left free
during smoothing, the values of these parameters fluctuated uncontrollably,
masking the contribution of the remaining parameters in the model
and, thus, achieving an almost perfect (but meaningless) smoothing
of the experimental data. This is an indication that the B4 model
we used may be missing one or more relevant components, such as additional
currents and compartments (see below for further analysis of this
point). We did not observe this effect in the cases examined in the
previous sections, where simulated data was used, because the models
responsible for the generation of this data were, by definition, precisely
known. 

Our results from this set of simulations are illustrated in Fig. 12.
Simultaneous smoothing of all four data sets was again accomplished
with high fidelity, as illustrated in Figs. 12Ai-iv. The artificial
evolution of the expectations of the conductances for the transient
sodium, persistent potassium and transient potassium currents, as
well as of some of the kinetic parameters that were estimated in the
model is illustrated in Figs. 12Bi-iii. The distributions of all inferred
parameters (normalized after replacing $x_{true}$ in Eq. \ref{eq:normrule}
with $\sum_{k=1}^{K}\bar{x}_{k}$, for each tested parameter) are
also illustrated in Fig. 12C. The inferred expectations of all parameters
are given in Table 3. 

In order to examine the predictive value of the model given the estimated
parameter expectations in Table 3, we compared its activity to that
of the biological B4 neuron, when both were injected with a $30s$-long
random current consisting of a sequence of current pulses with amplitude
ranging from $-4nA$ to $+4nA$ and duration from $1ms$ to $256ms$.
Our results from this simulation are illustrated in Fig. 13. We observed
that the overall pattern of activity of the model was similar to that
of the biological neuron (Fig. 13A). Whilst the model overall generated
more action potentials, some individual spikes were absent in the
simulated data. A more detailed examination of our data revealed specific
differences between the biological and model neurons, which explain
the differences in the overall activity between the two (Fig. 13B,
C). The spike shape of the model neuron was quite similar to that
of its biological counterpart (Fig. 13Bi), including spike threshold,
peak, trough and height (i.e. trough-to-peak amplitude; Fig. 13Biii),
but the simulated spike had a slightly longer duration than the biological
one (half-width: $1.9ms$ vs $1.5ms$; Fig. 13Bii). 

In a second set of experiments, both the biological and model neurons
were injected with $1s$-long current pulses ranging from $-4nA$
to $+4nA$ and their current-voltage (IV) and current-frequency (IF)
relations were constructed (Fig. 13C). The IV plot showed some non-linear
behavior in response to negative current pulses in the experimental
data (probably due to the presence of a residual $I_{h}$ current),
which was not present in the simulations (Fig. 13Ci). As a result,
the slope of the part of the IV curve corresponding to $0mV$ was
more shallow in the simulations than in the experimental data. Moreover,
the rheobase was lower in the experimental data than in the model,
but the slope of the IF curve was steeper in the simulated data, which
resulted in higher firing rates for the model at injected currents
larger than approximately $3nA$ (Fig. 13Cii). This feature can account
for the overall level of spiking in the model neuron when compared
to the biological one (Fig. 13A). 

Overall, this analysis illustrates that the assumed B4 model did not
capture all the aspects of the real neuron. However, this does not
mean that our estimation method is flawed. It just shows that the
model is actually missing some relevant components, such as additional
ionic currents or compartments, which would be necessary for approximating
more accurately the spatial structure and biophysical properties of
the biological neuron. In the first part of the manuscript we have
demonstrated that if the underlying model is complete, then our method
produced accurate estimates of the true parameter values, given sufficient
informative priors. Thus, it is safe to assume that the observed differences
between the biological and model neurons can be minimized, if the
fixed-lag smoother is applied on a more complex model of the B4 motoneuron. 

In summary, we used the fixed-lag smoother to estimate the unknown
parameters in a single-compartment model of an invertebrate motoneuron
based on actual electrophysiological data. The model, although a simplification
of the actual biological system, was still quite complex containing
a number of non-linearly interacting components and a total of $17$
unknown parameters. By using the methodologies outlined in the previous
sections, we managed to estimate the values of these parameters, such
that the resulting model mimicked with satisfactory accuracy the overall
activity of its biological counterpart. Furthermore, we demonstrated
the flexibility of the fixed-lag smoother by showing how it can be
used to process simultaneously multiple data sets, given an appropriate
formulation of the extended state vector.

\section*{Discussion}

Parameter estimation in conductance-based neuron models traditionally
involves a global optimization algorithm (for example, an evolutionary
algorithm), usually in combination with a local search method (such
as gradient descent), in order to find combinations of model parameters
that minimize a pre-defined cost function. In this paper, we have
addressed the problem of parameter estimation in Hodgkin-Huxley-type
models of single neurons from a different perspective. By adopting
a hidden-dynamical-systems formalism and expressing parameter estimation
as an inference problem in these systems, we made possible the application
of a range of well-established inference methods from the field of
Computational Statistics. Although it is usually assumed that the
kinetics of ionic currents in a conductance-based model are known
\textit{a priori}, here we assumed that this was not the case and,
typically, we estimated kinetic parameters, along with the maximal
conductances and reversal potentials of ionic currents in the models
we examined.

The particular method we used was Kitagawa's self-organizing state-space
model, which was implemented as a fixed-lag smoother. The smoother
was combined with an adaptive algorithm for sampling new sets of parameters
akin to the Covariance Matrix Adaptation Evolution Strategy. Alternatively,
we could have approximated the smoother distribution (Eq. \ref{eq:smoother})
with a two-pass algorithm, consisting of a forward filter followed
by a backward smoothing phase, which would make use of the precomputed
filter\cite{Cappe:2007}. This would require storing the filter for
the whole duration of the smoothed data, which in turn would have
very high memory requirements when large numbers of particles or high-dimensional
problems are considered. In contrast, the fixed-lag smoother has the
advantage that only the particles up to $L$ time steps in the past
need to be stored, which is less demanding in memory size and computationally
more efficient. Moreover, the fixed-lag smoother, being a single-pass
algorithm, was more natural to use in the context of on-line parameter
estimation. 

The applicability of the algorithm was demonstrated on a number of
conductance-based models using noisy simulated or actual electrophysiological
data. In a recent study, it was found that increasing observation
noise led to an increase in the variance of parameter estimates and
a decrease in the rate of convergence of the algorithm\cite{Huys:2009qa}.
Similarly, we observed that at high levels of observation noise, although
the algorithm remained functional, its accuracy and precision were
reduced (Fig. 5). It is emphasized that, at a particular level of
observation noise, the outcome of the algorithm is an approximation
of the posterior distributions of hidden states and unknown parameters
in the model, given the available experimental data and prior information.
In general, these approximate posteriors provide an overview of the
structure of the parameter space and they potentially have multiple
modes (or local optima). By taking advantage of the adaptive nature
of the fixed-lag smoother (and, in particular, by controlling the
scaling factor that determines the width of the proposal distribution
in Eq. \ref{eq:updtheta}), we managed to reduce the variance of these
posteriors and, in the limit case, we could force the algorithm to
converge to a single optimal point (belonging to the support of the
parameter posteriors), which could subsequently be used in predictive
simulations (e.g. see Figs. 7D and 11A). Unlike the study in \cite{Huys:2009qa},
we did not observe any significant reduction in the rate of convergence
of the algorithm at high levels of observation noise, which was attributed
to a ceiling effect due to the large number of particles we used in
our simulations (typically, $100\times D$, where $D$ was the dimensionality
of the estimation problem; see Figs. 3B and 4C). Thus, we cannot exclude
observing such a reduction in the rate of convergence, if a smaller
number of particles is used and/or problems of higher dimension are
examined. Furthermore, the proposed method requires only a single
forward pass of the experimental data, instead of multiple passes,
as in the case of off-line estimation methods, including the Expectation
Maximization (EM) algorithm. On the other hand, this means that, in
general, the proposed algorithm requires processing longer data time
series in order to converge. In addition, unlike off-line estimation
methods, it does not take into account the complete data trace at
each iteration, but at most $L$ past data points (but, also, see
\cite{Leeuwen:2010} for a partial ``remedy'' of this situation).
In principle, it would be possible to combine previous work on parameter
estimation (e.g. \cite{Haufler:2007, Huys:2006fk}) within an EM inference
framework in order to estimate various types of parameters (including
maximal conductances and channel kinetics) in conductance-based neuron
models. This could be an interesting topic for further research.

Our main conclusion was that, using this algorithm and a set of low-dimensional
experimental data (typically, one or more traces of membrane potential
activity), it was possible to fit complex compartmental models to
this data with high fidelity and, simultaneously, estimate the hidden
dynamic states and optimal values of a large number of parameters
in these models. Based on simulation experiments using simulated data,
we found that the estimated optimal parameter values and hidden states
were close to their true counterparts, as long as sufficient prior
information was made available to the algorithm. This information
took the form of knowledge of the values of particular parameters
(for example, the passive properties of the membrane) or of relatively
narrow ranges of permissible parameter values. Such prior information
could have included the kinetics of the ion currents that flow through
the membrane or the spatial distribution of various parameter values
along different neuron compartments (e.g. the ratio of maximal conductance
A between compartment 1 and compartment 2). In real-life situations,
such information may become available through current- or voltage-clamp
experiments. For example, the passive properties in the B4 model (membrane
capacitance, leakage maximal conductance and reversal potential) were
inferred from current-clamp data and, thus, they were fixed during
the subsequent smoothing phase. 

It has been demonstrated that this requirement for prior information
may be relaxed, if the data set used as input was sufficiently variable
to tease apart the relative contribution of different parameters in
a model\cite{Hobbs:2008oq}. A well-established result in conductance-based
modeling is that the same pattern of electrical activity may be produced
by different parameter configurations of the same model\cite{Goldman:2001ve, Golowasch:2002qf, Prinz:2003nx, Prinz:2004cr}.
This implies that it is impossible to identify, during the course
of an optimization procedure, a unique set of parameters using just
this single pattern of activity as input to the method. For example,
as we observed in the case of the B4 model, the posteriors of the
estimated parameters may be characterized by multiple modes (i.e.
local optima) or quite large variances, which makes identification
of a unique set of optimal parameter values for use in predictive
simulations rather difficult (Supplementary Fig. S7). A more variable
data set would be necessary in order to constrain the model under
study, thus forcing the optimization process to converge towards a
unique solution. It should be noted that this conclusion was reached
by treating as unknown only the maximal conductances in a conductance-based
model\cite{Hobbs:2008oq}. Although it is reasonable to assume that
this holds true when the kinetics of ion channels are also treated
as unknown, it still needs to be demonstrated whether the generation
of a data set sufficiently variable to constrain both the maximal
conductances and kinetics of ion channels in a complex conductance-based
model is practical or even feasible. A more pragmatic approach would
be to rely on a mixture of prior information and one or more sufficiently
variable electrophysiological recordings as input to the optimization
algorithm. It was shown in this study that both the injection of prior
information (in the form mentioned above) and the simultaneous assimilation
of multiple data sets is straightforward using the proposed algorithm. 

It is important to notice that, unlike more traditional approaches,
explicitly defining a cost or fitness function was not required by
the fixed-lag smoother. Given the fact that the efficiency of any
optimizer can be seriously impeded by a poorly designed cost function,
bypassing the need to define such a function may be viewed as an advantage
of the proposed method. As in previous studies\cite{Haufler:2007, Huys:2006fk},
here lies the implicit assumption that by fitting (or smoothing) with
high fidelity the raw experimental data (for example, one or more
recordings of the membrane potential), the estimated model would capture
a whole range of features embedded in this data, such as the current-frequency
response of the neuron. Although this is a reasonable assumption,
we found that it did not hold completely true, when our knowledge
of the form of the underlying model was not exact, as in the case
of the B4 neuron. In this case, although we could achieve a very good
smoothing of the experimental data, subsequent predictive simulations
using the inferred model parameters revealed discrepancies between
simulation output and experimental data. It is likely that these discrepancies
will be minimized, if important missing components are added to the
model, such as additional ionic currents or, importantly, an approximation
of the spatial structure of the biological neuron. 

An important outcome of this study was to demonstrate the intimate
relation between the self-organizing state-space model and evolutionary
algorithms. When used for parameter estimation, the self-organizing
state-space model undergoes at each iteration a process of new particle
(individual) generation (mutation/recombination) and resampling (selection
and multiplication), which parallels similar processes in evolutionary
algorithms. At the root of this parallelism is the fact that we need
to impose an artificial evolution on model parameters as part of the
formulation of the self-organizing state-space model (see Methods),
thus providing a unique opportunity to merge the two classes of algorithms.
Here, we decided to combine the self-organizing state-space model
with an adaptive algorithm similar to the Covariance Matrix Adaptation
Evolution Strategy\cite{Igel:2007} and by following this adaptive
strategy, we managed to achieve a dramatic reduction in the variance
of parameter estimates. However, this choice is by no means exclusive
and other evolutionary algorithms may be chosen instead, e.g. the
Differential Evolution algorithm\cite{Price:2005}. This is a topic
open to further exploration. Notice that, similarly to Evolutionary
Algorithms, the proposed method has, in principle, the ability to
estimate the possibly multi-modal posterior distribution of the unknown
parameters in the examined model, i.e. it is a global estimation method
(for example, see Fig. 6C, 11B and S7). At each iteration, the algorithm
retains a population of particles, which are characterized by a degree
of variability and, thus, give the algorithm the opportunity to randomly
explore a wide range of the parameter space, spending on average more
time in the vicinity of optimal regions. By imposing narrow prior
constraints on some of the unknown parameters, we are effectively
reducing the dimensionality of the problem and we force the algorithm
to converge towards a particular optimum, which can be later used
in predictive simulations. 

A point of potential improvement concerns our choice of the proposal
density, $q(\mathbf{z}_{k}|\mathbf{z}_{k-1},\mathbf{y}_{k})$. Here,
we made the common and straightforward choice to use the transition
density $p(\mathbf{z}_{k}|\mathbf{z}_{k-1})$ as our proposal. However,
the modeler is free to make other choices. For example, a recent study
demonstrated that the efficiency of particle filters can be significantly
increased by conditioning the proposal density on future observations\cite{Leeuwen:2010}.

An important practical aspect of the proposed algorithm was its high
computational cost. This cost increased as a function of the number
$N$ of particles used during smoothing, the length of the fixed smoothing
lag $L$, the complexity of the model and the number of unknown parameters
in the model. Our simulations on an Intel dual-core i5 processor with
four gigabytes of memory took from a few minutes to more than 12 hours
to complete. An emerging trend in Scientific Computing is the use
of modern massively parallel Graphics Processing Units (GPUs) in order
to accelerate general purpose computations, as those presented in
this paper. The utility of this approach in achieving significant
accelerations of Monte Carlo simulations has been recently demonstrated\cite{Lee:2010uq}
and it has even been applied recently on parameter estimation problems
in conductance-based models of single neurons\cite{Quinn:2011kx}.
Preliminary results using a GPU-accelerated version of the fixed-lag
smoother (data not shown) have indeed demonstrated reduced simulation
times, but the accelerations we observed were not as dramatic as those
reported in the literature\cite{Lee:2010uq,Quinn:2011kx}. This can
always be attributed to the fact that our implementation of the algorithm
was not optimized. On the other hand, we observed significant accelerations
in our simulations involving the serial implementation of the fixed-lag
smoother, just by switching from an open-source compiler (GNU) to
a commercial one (Intel), which presumably emitted better optimized
machine code for the underlying hardware. Nevertheless, the use of
GPUs for general purpose computing is becoming common and it is likely
to become quite popular with the advent of cheaper hardware and, importantly,
more flexible and programmer-friendly Application Programming Interfaces
(APIs). 

Overall, our results point towards a generic four-stage heuristic
for parameter estimation in conductance-based models of single neurons:
(a) First, the general structure of the model is decided, such as
the number of ionic currents and compartments it should include. (b)
Second, prior information is exploited in order to fix as many parameters
as possible in the model and tightly constrain the remaining ones.
For example, the capacitance, reversal potentials and leakage conductance
in the model may be fixed to values estimated from current-clamp data.
By further exploiting current-- and voltage-clamp data, narrow constraints
may be imposed on the remaining free (e.g. kinetic) parameters in
the model. (c) At a third stage, more precise parameter value distributions
are estimated by applying the fixed-lag smoother on current-clamp
data, such as one or more recordings of the electrical activity of
the membrane induced by random current injections. (d) Finally, the
predictive value of the model is assessed through comparison to independent
data sets and the model is modified, if necessary. It is important
to notice that the techniques outlined in this paper are applicable
on a wide range of research domains and that they provide a disciplined
way to merge complex stochastic dynamic models, noisy data and prior
information under a common inference framework. 

In conclusion, the class of statistical estimation methods, which
the algorithm presented in this paper belongs to, in combination with
Monte Carlo approximation techniques are particularly suitable to
address high-dimensional inference problems in a disciplined manner.
This makes them potentially useful tools at the disposal of biophysical
modelers of neurons and neural networks and it is predicted that these
methodologies will become more popular in the future among this research
community.

\section*{Acknowledgments}

This study was supported by the Future and Emerging Technologies (FET)
European Commission program BION (number 213219).

\bibliography{manuscript}

\clearpage

\section*{Figures}

\begin{figure}[!ht]
\includegraphics[width=4in]{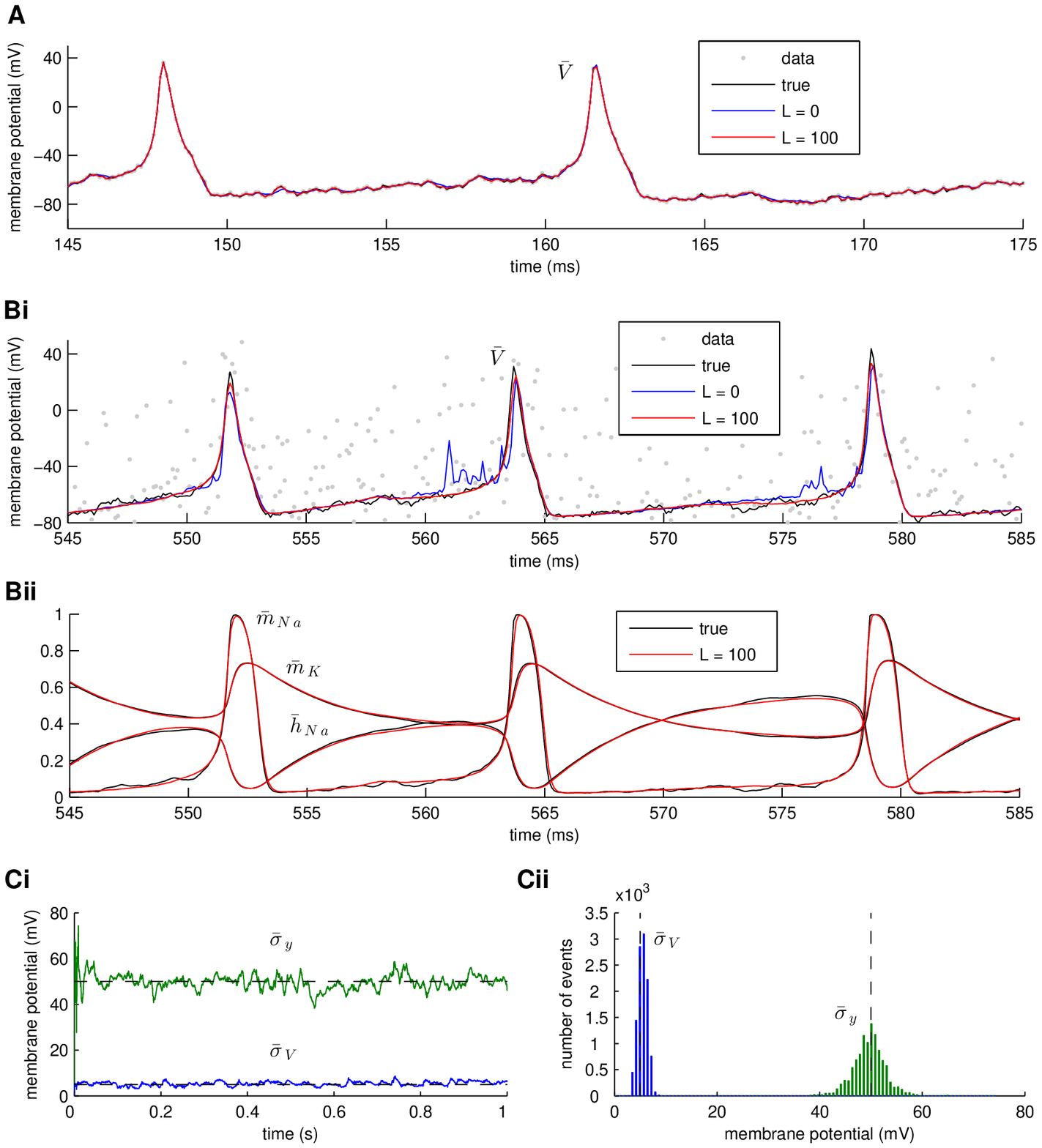}
\caption{\textbf{Simultaneous estimation of hidden states, intrinsic and observation
noise}. Estimation was based on a simulated recording of membrane
potential with duration $1s$. For clarity, only $30ms$ of activity
are shown in A and Bi,ii. (\textbf{A}) Smoothing of the membrane potential
(the observed variable), when observation noise was low ($\sigma_{y}=0.5mV$).
High-fidelity smoothing was achieved for either small ($L=0$) or
large ($L=100$) values of the fixed smoothing lag $L$. Simulated
and smoothed data are difficult to distinguish due to their overlap.
(\textbf{Bi}) Smoothing of the membrane potential at high levels of
observation noise ($\sigma_{y}=50mV$). A large value of the smoothing
lag ($L=100$) was required for high-fidelity smoothing. (\textbf{Bii})
Inference of the unobserved activation ($m_{Na}$, $m_{K}$) and inactivation
($h_{Na}$) variables for sodium and potassium currents as functions
of time, during smoothing of the data shown in Bi for $L=100$. (\textbf{Ci})
Inference of the standard deviations for the intrinsic and observation
noise ($\sigma_{V}$ and $\sigma_{y}$, respectively) during smoothing
of the data shown in Bi for $L=100$. Dashed lines indicate the true
values of $\sigma_{V}$ and $\sigma_{y}$. (\textbf{Cii}) Histograms
of the time series for $\sigma_{V}$ and $\sigma_{y}$ in Ci. Again,
dashed lines indicate the true values of the corresponding parameters.
At this stage, maximal conductances, reversal potentials and kinetic
parameters in the model were assumed known. The number of particles
was $N=700$. Also, $a=b=c=0$. The scaling factors in Eq. \ref{eq:updtheta}
were all considered equal to $1$.}
\end{figure}

\begin{figure}[!ht]
\includegraphics[width=4in]{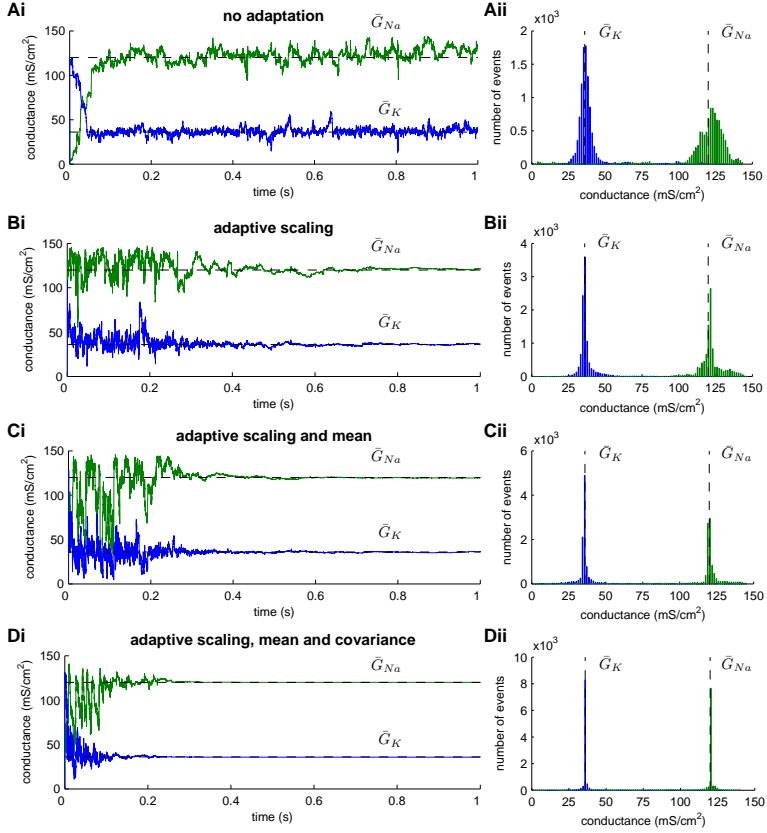}
\caption{\textbf{The effect of adaptive parameter sampling on the variance
of parameter estimates.} Merging the fixed-lag smoother with an adaptive
sampling algorithm akin to the Covariance Matrix Adaptation Evolution
Strategy reduced significantly the variance of parameter estimates.
At this stage, the maximal conductances for the sodium ($G_{Na}$)
and potassium ($G_{K}$) currents were assumed unknown. Estimation
was based on a simulated recording of membrane potential with duration
$1s$ and $\sigma_{V}=\sigma_{y}=1mV$. (\textbf{A}) Inference of
$G_{Na}$ and $G_{K}$ during smoothing, when new parameter samples
were drawn from a non-adaptive multi-variate normal distribution (Eq.
\ref{eq:bivnorm}). Dashed lines indicate the true parameter values.
(\textbf{B}) Inference of $G_{Na}$ and $G_{K}$ during smoothing,
when new samples were drawn from a multi-variate normal distribution
(Eq. \ref{eq:updtheta}) with an adaptive scaling factor $s$ ($c=0.01$
in Eq. \ref{eq:lognorm}). (\textbf{C}) Inference of $G_{Na}$ and
$G_{K}$ during smoothing, when new samples were drawn from a multi-variate
normal distribution (Eq. \ref{eq:updtheta}) with adaptive scaling
(as in B) and mean ($a=0.01$ in Eq. \ref{eq:meantheta}). (\textbf{D})
Inference of $G_{Na}$ and $G_{K}$ during smoothing, when new samples
were drawn from a multi-variate normal distribution with adaptive
scaling (as in B), mean (as in C) and covariance ($b=0.01$ in Eq.
\ref{eq:covtheta}). The histograms in the right plots were constructed
from the time series in the left plots. Membrane potential, activation
and inactivation variables, intrinsic and observation noise were also
subject to estimation, as in Fig. 1. Smoothing lag and number of particles
were $L=100$ and $N=900$, respectively. The prior interval of the
scaling factors $s_{k}^{(j)}$ was $[0,10]$.}

\end{figure}

\begin{figure}[!ht]
\includegraphics[width=4in]{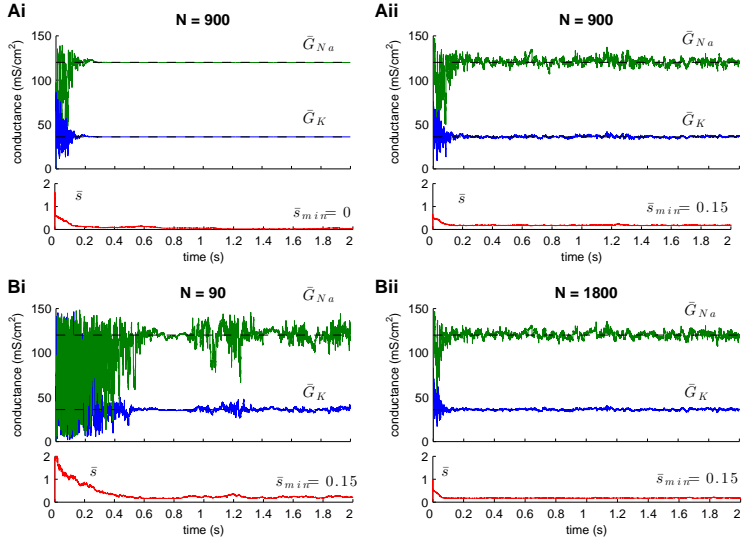}
\caption{\textbf{The effect of the size of the scaling factor $s$ and the
number of particles $N$ on the variance of the estimates.} Large
minimal values of $s$ and small values of $N$ imply large variance
of the estimates. (\textbf{A}) Resampling of particles (see Methods)
implies adaptation of (among others) the scaling factors $s_{k}^{(j)}$,
which gradually approach the lower bound of their prior interval (red
lines in Ai,ii). A prior interval with zero lower bound (i.e. $[0,2]$)
leads to estimates with negligible variance (Ai). A prior interval
with relatively large lower bound (e.g. $[0.15,2]$) leads to estimates
with non-zero variance (Aii). Notice that the expectation $\bar{s}$
in Ai does not actually take the value $0$ (instead it becomes approximately
equal to $0.01$). (\textbf{B}) A small number of particles (Bi, $N=90$)
implies estimates with large variance (compare to Bii, $N=1800$).
Notice that the difference between Aii ($N=900$) and Bii ($N=1800$)
is negligible, implying the presence of a ceiling effect, when the
number of particles becomes very large. In these simulations, $L=100$
and $a=b=c=0.01$.}

\end{figure}

\begin{figure}[!ht]
\includegraphics[width=4in]{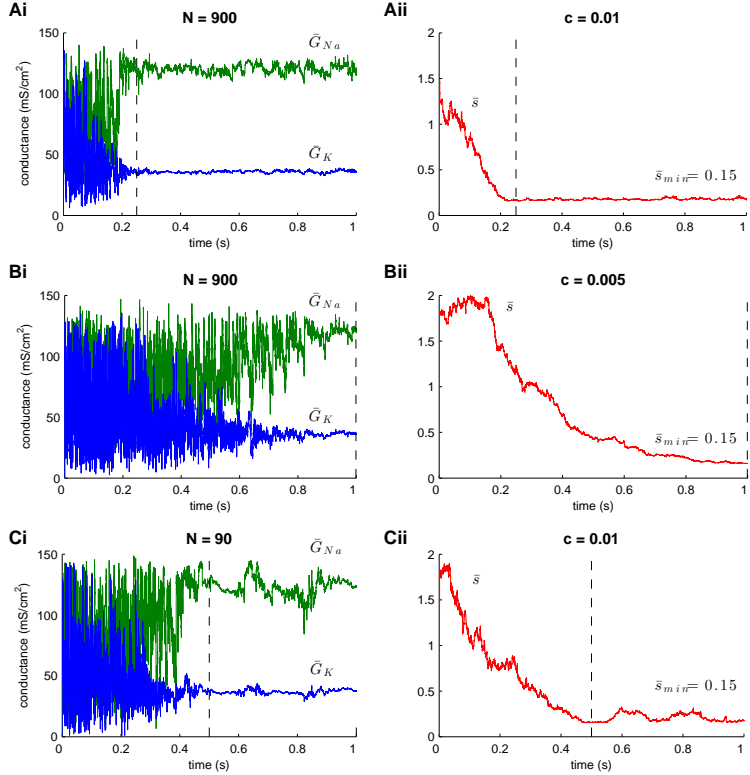}
\caption{\textbf{The effect of adaptation of the scaling factor $s$ and the
number of particles $N$ on the speed of convergence.} A slow rate
of adaptation for $s$ and a small number of particles $N$ imply
slow convergence of the algorithm. The rate at which $s_{k}^{(j)}$
adapts depends on the parameter $c$ in Eq. \ref{eq:lognorm}. Reducing
$c$ in half results in a significant decrease in the rate of convergence
(compare A to B). Also, reducing the number of particles by a factor
of $10$ slows down the speed of convergence (compare A to C), but
not as much as when parameter $c$ was adjusted. The plots on the
right illustrate the profile of $\bar{s}$ associated with the estimation
of the parameters on the left plots. In these simulations, $L=100$,
$a=b=0.01$ and the prior interval for the scaling factors $s_{k}^{(j)}$
was $[0.15,2]$.}

\end{figure}

\begin{figure}[!ht]
\includegraphics[width=4in]{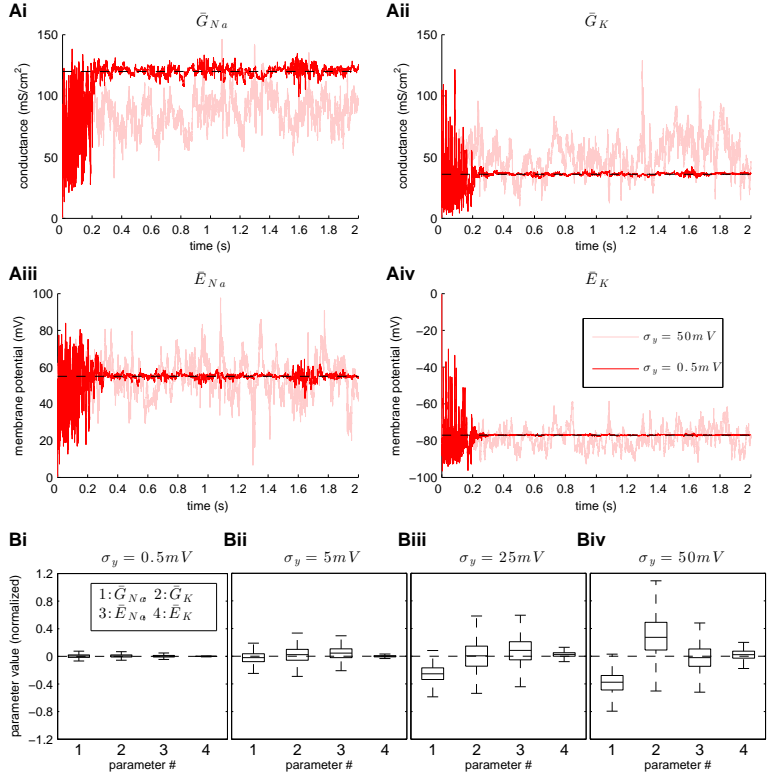}
\caption{\textbf{The effect of observation noise on the accuracy and precision
of parameter estimates.} Increasing observation noise decreases the
accuracy and precision of the fixed-lag smoother. At this stage, the
reversal potentials for the sodium and potassium currents ($E_{Na}$
and $E_{K}$, respectively) were also considered unknown. Estimation
was based on a simulated recording of membrane potential with duration
2s. The noise parameters were $\sigma_{V}=1mV$ and $\sigma_{y}=0.5mV$,
$5mV$, $25mV$ or $50mV$. (\textbf{A}) Inference of $G_{Na}$, $G_{K}$,
$E_{Na}$ and $E_{K}$ during smoothing. The accuracy of the estimates
decreases and their variance increases with increasing observation
noise. (\textbf{B}) The box plot of the time series in A for $t\ge1s$.
Data were first normalized according to Eq. \ref{eq:normrule}. The
reduction in the accuracy and precision at higher levels of observation
noise were more prominent in the case of the maximal conductances
($G_{Na}$ and $G_{K}$) and less prominent in the case of reversal
potentials ($E_{Na}$ and, particularly $E_{K}$). The membrane potential,
activation and inactivation variables, intrinsic and observation noise
were also subject to estimation, as in Fig. 1. In these simulations,
$L=100$, $N=1100$, $a=b=c=0.01$ and the prior interval of $s_{k}^{(j)}$
was $[0.15,10]$.}

\end{figure}

\begin{figure}[!ht]
\includegraphics[width=4in]{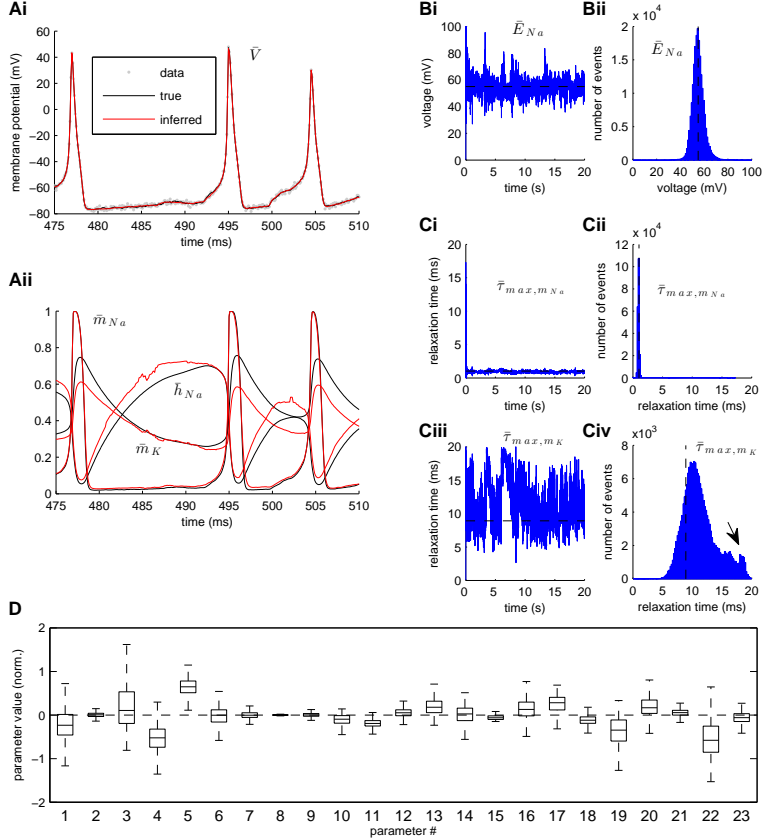}
\caption{\textbf{Estimation of all parameters in a single-compartment conductance-based
model using the fixed-lag smoother.} Estimation was based on a simulated
recording of the membrane potential with duration $20s$. Noise parameters
were $\sigma_{V}=\sigma_{y}=1mV$. For clarity, only $35ms$ of activity
are illustrated in Ai,ii. (\textbf{A}) Smoothing of the membrane potential
(Ai) and the unobserved activation and inactivation variables for
the sodium and potassium currents (Aii). (\textbf{B, C}) Estimated
posteriors for $E_{Na}$ (B), $\tau_{max,m_{Na}}$ (Ci,ii) and $\tau_{max,m_{K}}$
(Ciii,iv). The histograms on the right were constructed form the data
on left. (\textbf{D}) Box plot of the $23$ estimated parameter posteriors
in the model. These included the standard deviations of intrinsic
and observation noise, maximal conductances, reversal potentials and
kinetics of all currents in the model (see Table 1). The estimates
were first normalized according to Eq. \ref{eq:lognorm}. Parameter
identification numbers are as in Table 1. In these simulations, $L=100$,
$N=2800$, $a=b=c=0.01$ and the prior interval for $s_{k}^{(j)}$
was $[0.15,10]$. }

\end{figure}

\begin{figure}[!ht]
\includegraphics[width=4in]{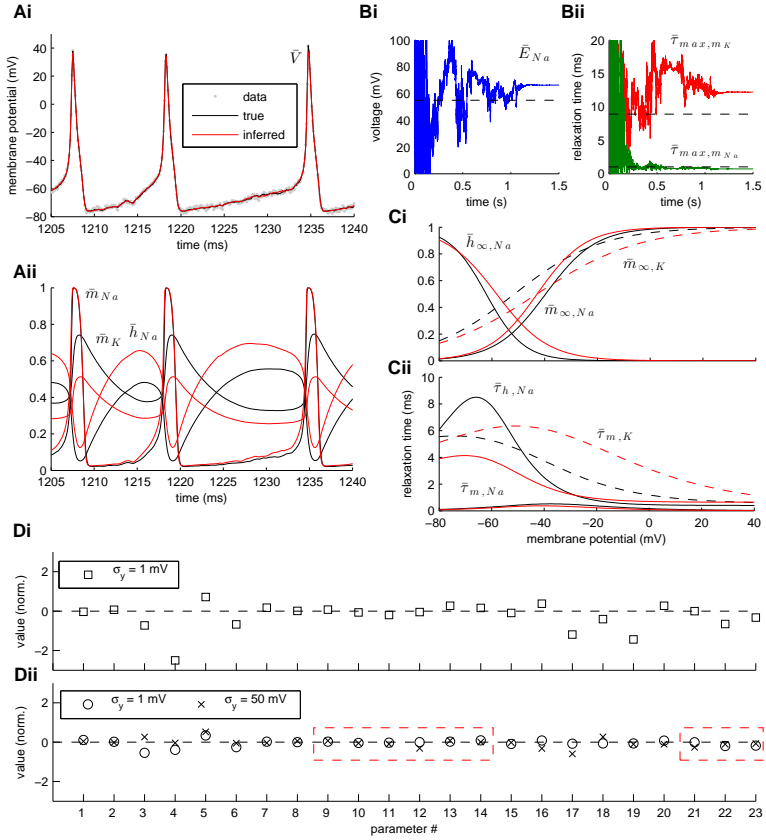}
\caption{\textbf{The effect of prior parameter intervals on the accuracy of
the fixed-lag smoother. }Estimation was based on a simulated recording
of the membrane potential with duration $2s$. Noise parameters were
$\sigma_{V}=\sigma_{y}=1mV$. For clarity, only $35ms$ of activity
are illustrated in Ai,ii. Unlike Fig. 6, the prior interval for the
scaling factors $s_{k}^{(j)}$ was now assumed equal to $[0,10]$.
(\textbf{A}) Smoothing of the membrane potential (Ai) and the unobserved
activation and inactivation variables for the sodium and potassium
currents (Aii). (\textbf{B}) Estimates for parameters $E_{Na}$ (Bi),
$\tau_{max,m_{Na}}$ and $\tau_{max,m_{K}}$ (Bii). Convergence to
an optimal parameter vector was achieved after approximately $1.5s$
of activity. Notice that this optimal parameter vector falls within
the support of the corresponding parameter posteriors (see Figs. 6Bii,
6Cii and 6Civ). (\textbf{C) }Inferred steady states (Ci) and relaxation
times (Cii) for the activation and inactivation variables of sodium
and potassium currents (red lines) against their true counterparts
(black lines). (\textbf{D}) Inferred parameter values when broad (Di)
or narrow (Dii) prior intervals were used for the parameters controlling
the kinetics of sodium and potassium ionic currents (see Table I).
Plots A, B and C correspond to plot Di. In Dii, we also illustrate
the estimated parameter values when very noisy data were used (see
also Supplementary Fig. S3). In these simulations, $L=100$, $N=2800$
and $a=b=c=0.01$.}

\end{figure}

\begin{figure}[!ht]
\includegraphics[width=4in]{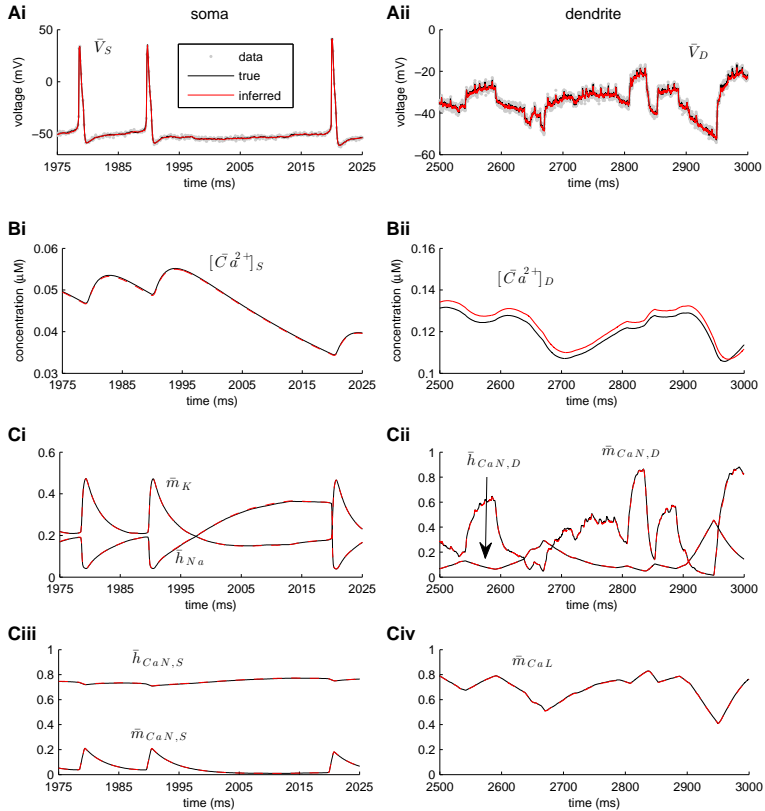}
\caption{\textbf{Simultaneous estimation of hidden model states (including
intracellular calcium concentrations) and maximal conductances in
a two-compartment model of a vertebrate motoneuron (I).} Estimation
was based on two $3s$-long simulated recordings of the membrane potential,
each recorded simultaneously from the soma and the dendritic compartment.
Only part of the recorded activity is illustrated in A, B and C for
clarity. Notice the different time scales between the right and left
panels. (\textbf{A}) High-fidelity smoothing of the membrane potential
at the soma (Ai) and the dendritic compartment (Aii). (\textbf{B})
Inference of the unobserved calcium concentrations at the soma (Bi)
and the dendrite (Bii). (\textbf{C}) Inference of the unobserved activation
and inactivation variables for the sodium and potassium currents (Ci)
and the N-type calcium current (Ciii) at the soma and the N-type (Cii)
and L-type (Civ) calcium currents at the dendritic compartment. Notice
the almost complete overlap between true (black lines) and inferred
(red lines) dynamic variables in Ci-iv. This was not surprising since
we assumed, at this stage, that the kinetics of all gated currents
were known. In these simulations, $L=100$, $N=2200$, $a=b=c=0.01$
and the prior interval for $s_{k}^{(j)}$ was $[0,10]$. }

\end{figure}

\begin{figure}[!ht]
\includegraphics[width=4in]{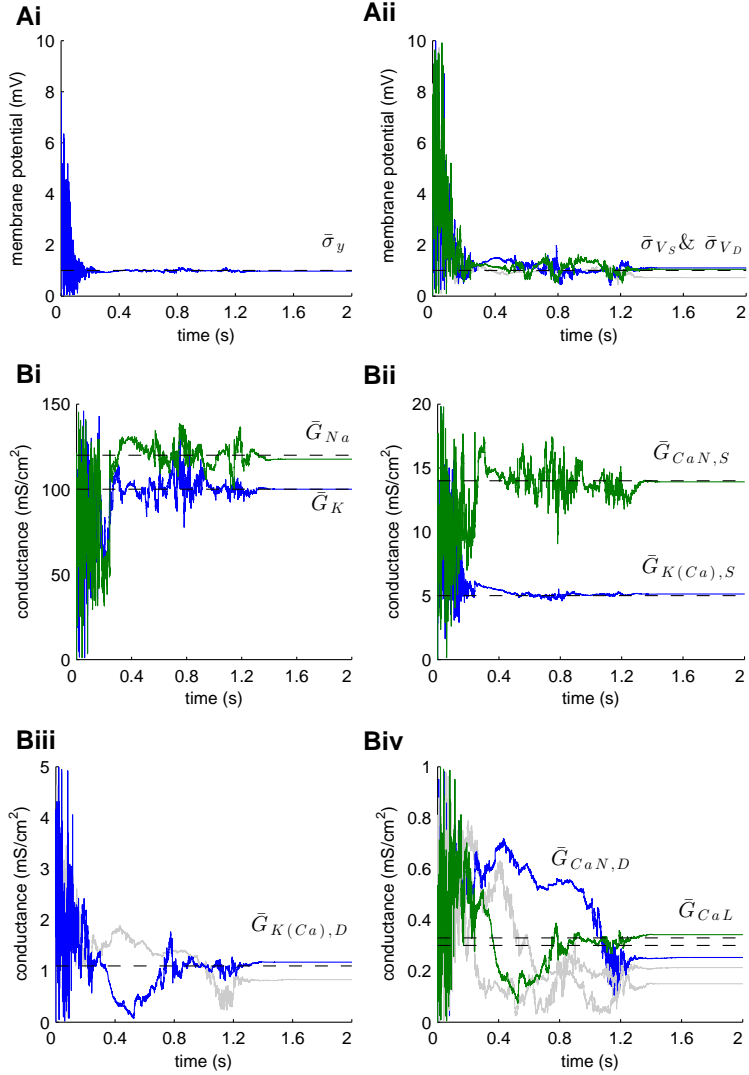}
\caption{\textbf{Simultaneous estimation of hidden model states (including
intracellular calcium concentrations) and maximal conductances in
a two-compartment model of a vertebrate motoneuron (II).} Inference
of maximal conductances and noise parameters during fixed-lag smoothing.
(\textbf{A}) The standard deviations of the observation (Ai) and the
intrinsic (Aii) noise at the soma and the dendrite. (\textbf{B}) Inferred
maximal conductances of the sodium and potassium currents at the soma
(Bi), of the N-type calcium current and the calcium-activated potassium
current at the soma (Bii), of the calcium-activated potassium current
at the dendrite (Biii) and of the N-type and L-type calcium currents
at the dendrite (Biv). In all cases, parameter expectations gradually
converged towards the true parameter values (dashed lines) after less
than $2s$. The grey lines in Aii, Biii and Biv correspond to estimated
parameters, when current was injected in the soma only. In these simulations,
$L=100$, $N=2200$, $a=b=c=0.01$ and the prior interval for $s_{k}^{(j)}$
was $[0,10]$. }

\end{figure}

\begin{figure}[!ht]
\includegraphics[width=4in]{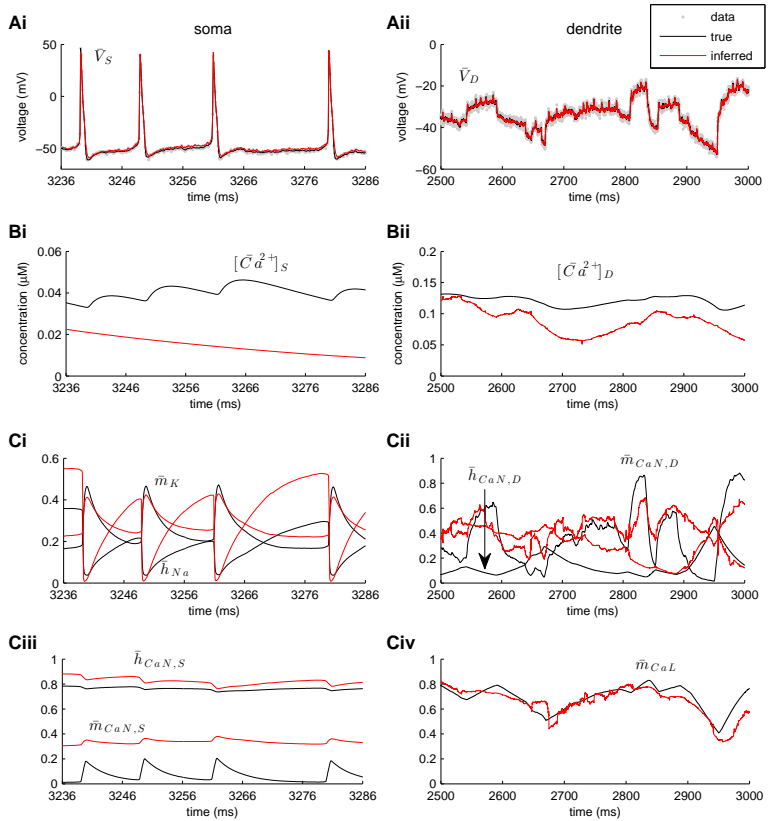}
\caption{\textbf{Simultaneous estimation of hidden model states, maximal conductances
and kinetic parameters in a two-compartment model of a vertebrate
motoneuron (I).} Estimation was based on two simulated $4s$-long
simultaneous recordings of the membrane potential from the soma and
dendritic compartment. Only part of this data is illustrated for clarity.
Notice the different time scales between the left and right panels.
(\textbf{A}) High-fidelity smoothing of the observed voltage at the
soma (Ai) and the dendrite (Aii). (\textbf{B}) Inference of unobserved
calcium concentrations at the soma (Bi) and dendritic compartment
(Bii). (\textbf{C}) Inference of the unobserved activation and inactivation
variables for all voltage-gated currents at the soma and the dendrite.
Since the kinetics of voltage-gated currents were assumed unknown,
the difference between true (black lines) and inferred (red lines)
dynamic variables was significant (compare to Fig. 8). The inferred
parameters are shown in Fig. 7Ai. In these simulations, $L=100$,
$N=8200$, $a=b=c=0.01$ and the prior interval for $s_{k}^{(j)}$
was $[0,10]$.}

\end{figure}

\begin{figure}[!ht]
\includegraphics[width=4in]{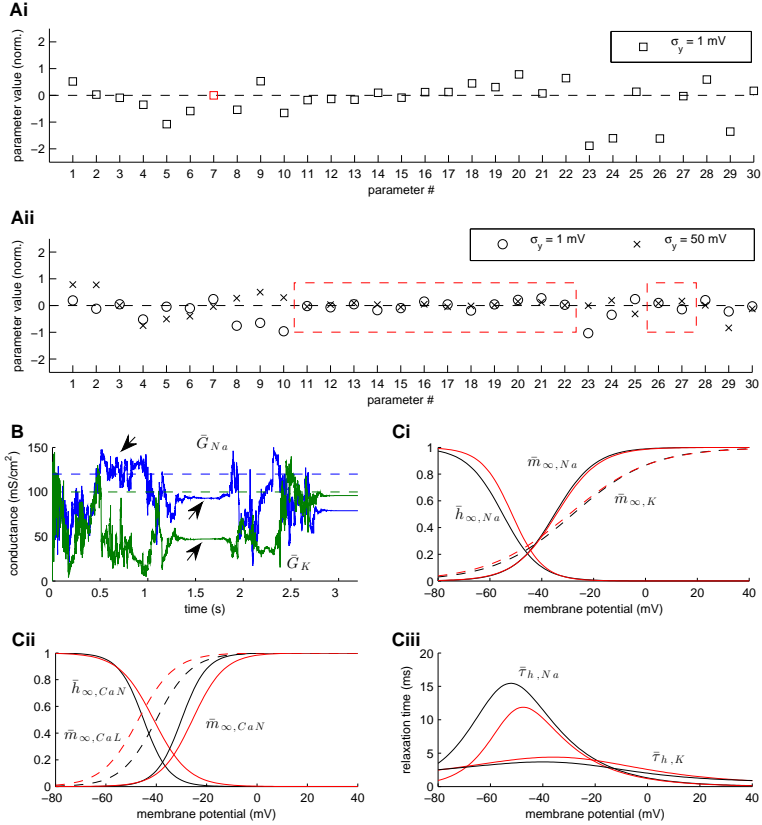}
\caption{\textbf{Simultaneous estimation of hidden model states, maximal conductances
and kinetic parameters in a two-compartment model of a vertebrate
motoneuron (II).} Inference of maximal conductances, noise and kinetic
parameters during smoothing. (\textbf{A}) Inferred parameters in the
model using broad or narrow prior intervals and high or low levels
of observation noise. Estimates were normalized according to Eq. \ref{eq:normrule}.
Parameter identification numbers are as in Table 2. The estimates
in Ai were obtained using broad prior intervals (see Table 2). The
maximal conductance $G_{CaN,S}$ (parameter \#7) converged to zero
and, for this reason, it is indicated with a red square. These estimates
correspond to the results shown in Fig. 10. Estimates in Aii were
obtained using narrow prior intervals for some of the parameters controlling
the kinetics of ionic currents (see red dashed boxes) at either low
($\sigma_{y}=1mV$) or high ($\sigma_{y}=50mV$) levels of observation
noise (see also Supplementary Figs. S4 and S5). (\textbf{B}) Inferred
maximal conductances for sodium ($G_{Na}$) and potassium ($G_{K}$)
when narrow prior intervals and low levels of observation noise were
used (circles in Aii). Notice the temporary convergence of the estimates
(arrows) before jumping away towards their final values. (\textbf{C})
True (black lines) and inferred (red lines) activation and inactivation
steady-states for the sodium and potassium currents (Ci) and the N-type
and L-type calcium currents (Cii) and for the relaxation times for
sodium inactivation and potassium activation (Ciii), when narrow prior
intervals and low levels of observation noise were used (circles in
Aii). In these simulations, $L=100$, $a=b=c=0.01$ and the prior
interval for $s_{k}^{(j)}$ was $[0,10]$. The number of particles
was $N=8200$ in Ai and $N=4100$ in Aii, B and C (see main text for
further comments).}
\end{figure}

\begin{figure}[!ht]
\includegraphics[width=4in]{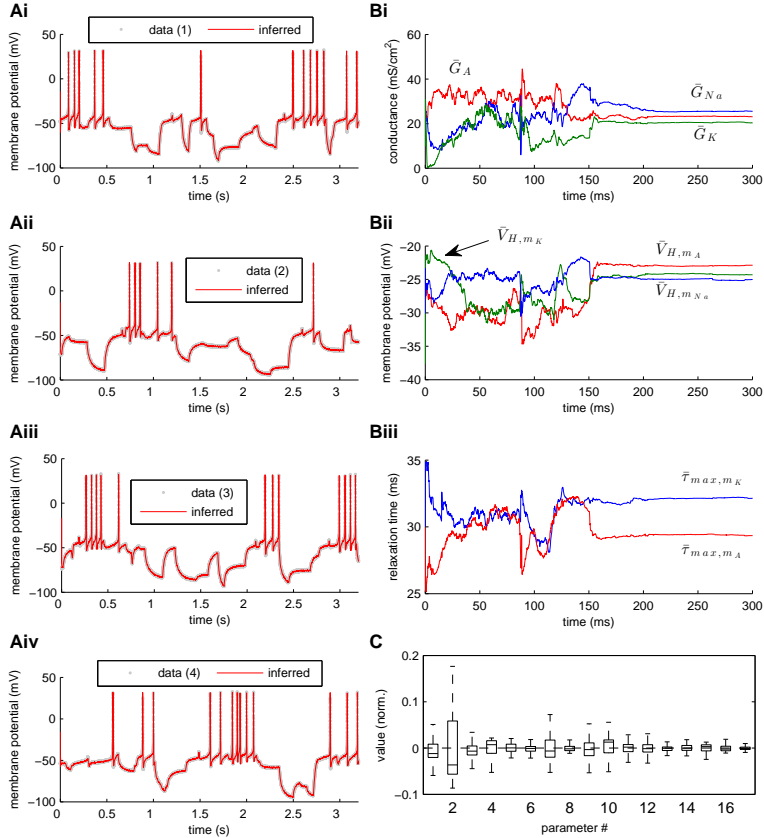}
\caption{\textbf{Parameter estimation in a model invertebrate motoneuron based
on actual electrophysiological data.} Estimation was based on four
independent $3.5s$-long recordings of the membrane potential from
the same B4 motoneuron. (\textbf{A}) Simultaneous, high-fidelity smoothing
of the four membrane potential recordings. (\textbf{B}) A total of
$17$ free parameters in the model were inferred during smoothing
(see Table 3), including the maximal conductances of the transient
sodium and potassium and persistent potassium currents (Bi), the half
steady-state activation values (Bii) and the relaxation times for
the activation of the potassium currents (Biii). The remaining inferred
parameters are not illustrated for clarity, but they follow a similar
convergence pattern. (\textbf{C}) Box plot of all inferred parameters
in the model. Parameter identification numbers are as in Table 3.
Estimates were normalized as explained in the main text (the non-normalized
mean parameter values are given in Table 3). In this simulation, $L=100$,
$N=3800$, $a=b=c=0.01$ and the prior interval for $s_{k}^{(j)}$
was $[0.2,0.5]$. }

\end{figure}

\begin{figure}[!ht]
\includegraphics[width=4in]{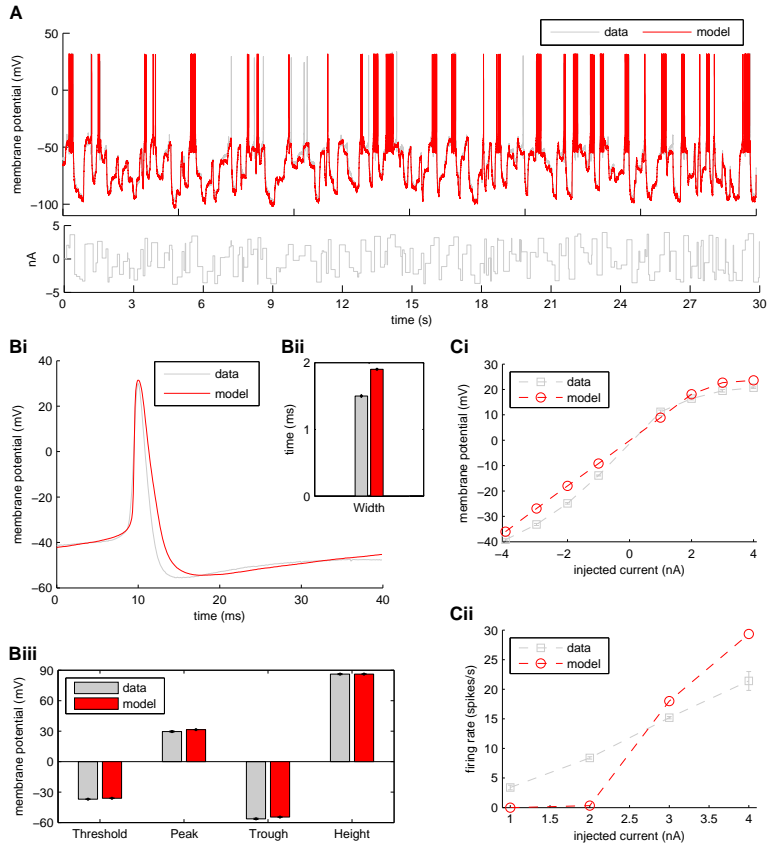}
\caption{\textbf{Comparison between B4 model activity and the biological neuron.}
(\textbf{A}) Response of the model and the biological B4 motoneuron
to a sequence of current steps with random amplitude and duration.
Current step amplitudes were from $-4nA$ to $+4nA$ and current step
durations from $1ms$ to $256ms$. Intrinsic and observation noise
in the model were $\sigma_{V}=0.3mV$ and $\sigma_{y}=1mV$, respectively.
(\textbf{B}) Comparison between model and biological B4 action potentials.
The width of the spikes was measured at half their peak amplitude.
(\textbf{C}) Current-Voltage (IV) and Current-Frequency (IF) relations
for the model and biological B4 neurons. In order to construct these
relations both the model and biological neurons were injected with
$1s$-long current pulses with amplitude between $-4nA$ and $+4nA$. }

\end{figure}

\clearpage

\section*{Tables}

\begin{table}[!ht]
\caption{\textbf{True and estimated values and prior intervals used during
smoothing for all parameters in the single-compartment conductance-based
model}}

\begin{tabular}{|c|l|l|r|r|r|r|}
\hline 
\# & Parameter & Unit & True Value & Estimated Value$^{1}$ & Lower Bound & Upper Bound\tabularnewline
\hline 
1 & $\sigma_{V}$ & $mV$ & $1.0$ & $1.0$ & $0.0$ & $10.0$\tabularnewline
2 & $\sigma_{y}$ & $mV$ & $1.0$ & $1.0$ & $0.0$1 & $10.0$\tabularnewline
3 & $G_{L}$ & $mS/cm^{2}$ & $0.3$ & $0.17$ & $0.0$ & $150.0$\tabularnewline
4 & $G_{Na}$ & $mS/cm^{2}$ & $120$.0 & 34.3 & $0.0$ & $150.0$\tabularnewline
5 & $G_{K}$ & $mS/cm^{2}$ & $36$.0 & $125.9$ & $0.0$ & $150.0$\tabularnewline
6 & $E_{L}$ & $mV$ & $-54.4$ & $-32.49$ & $-100.0$ & $0.0$\tabularnewline
7 & $E_{Na}$ & $mV$ & $55$.0 & $66.35$ & $0.0$ & $100.0$\tabularnewline
8 & $E_{K}$ & $mV$ & $-77$.0 & $-77.8$ & $-100.0$ & $0.0$\tabularnewline
9 & $V_{H,m_{Na}}$ & $mV$ & $-39.6$ & $-42.9$ & $-70.0$ \textbf{(-45.0)} & $-30.0$ \textbf{(-35.0)}\tabularnewline
10 & $V_{H,h_{Na}}$ & $mV$ & $-62.2$ & $-58.2$ & $-70.0$ \textbf{(-65.0)} & $-30.0$ \textbf{(-55.0)}\tabularnewline
11 & $V_{H,m_{K}}$ & $mV$ & $-51.5$ & $-43.0$ & $-70.0$ \textbf{(-55.0)} & $-30.0$ \textbf{(-45.0)}\tabularnewline
12 & $V_{S,m_{Na}}$ & $mV$ & $9.5$ & $9.0$ & $5.0$ \textbf{(5.0)} & $25.0$ \textbf{(10.0)}\tabularnewline
13 & $V_{S,h_{Na}}$ & $mV$ & $-7.1$ & $-9.7$ & $-25.0$ \textbf{(-10.0)} & $-5.0$ \textbf{(-5.0)}\tabularnewline
14 & $V_{S,m_{K}}$ & $mV$ & $16.4$ & $19.6$ & $5.0$ \textbf{(10.0)} & $25.0$ \textbf{(20.0)}\tabularnewline
15 & $\tau_{min,m_{Na}}$ & $ms$ & $0.0093$ & $0.009$ & $0.008$ & $1.0$\tabularnewline
16 & $\tau_{min,h_{Na}}$ & $ms$ & $0.4$ & $0.6$ & $0.01$ & $1.0$\tabularnewline
17 & $\tau_{min,m_{K}}$ & $ms$ & $0.5$ & $0.24$ & $0.01$ & $1.0$\tabularnewline
18 & $\tau_{max,m_{Na}}$ & $ms$ & $1.0$ & $0.7$ & $0.01$ & $20.0$\tabularnewline
19 & $\tau_{max,h_{Na}}$ & $ms$ & $16.1$ & $6.6$ & $0.01$ & $20.0$\tabularnewline
20 & $\tau_{max,m_{K}}$ & $ms$ & $8.9$ & $12.2$ & $0.01$ & $20.0$\tabularnewline
21 & $\delta_{m_{Na}}$ & - & $0.4$ & $0.4$ & $0.0$ \textbf{(0.0)} & $1.0$ \textbf{(0.5)}\tabularnewline
22 & $\delta_{h_{Na}}$ & - & $0.4$ & $0.2$ & $0.0$ \textbf{(0.0)} & $1.0$ \textbf{(0.5)}\tabularnewline
23 & $\delta_{m_{K}}$ & - & $0.8$ & $0.6$ & $0.0$ \textbf{(0.5)} & $1.0$ \textbf{(1.0)}\tabularnewline
\hline 
\end{tabular}

$^{1}$These parameter values were estimated when we used the broad
prior intervals (see Fig. 7Di)

$^{2}$Values in bold indicate the narrow prior intervals we used
for generating Fig. 7Dii (and Supplementary Fig. S3) 
\end{table}

\begin{table}[!ht]
\caption{\textbf{True and estimated values and prior intervals used during
smoothing for all parameters in the two-compartment conductance-based
model}}

\begin{tabular}{|c|l|l|r|r|r|r|}
\hline 
\# & Parameter & Unit & True Value & Estimated Value$^{1}$ & Lower Bound & Upper Bound\tabularnewline
\hline 
1 & $\sigma_{V_{S}}$ & $mV$ & $1.0$ & $2.1$ & $0.0$ & $10.0$\tabularnewline
2 & $\sigma_{V_{D}}$ & $mV$ & $1.0$ & $1.0$ & $0.0$ & $10.0$\tabularnewline
3 & $\sigma_{y}$ & $mV$ & $1.0$ & $0.9$ & $0.01$ & $10.0$\tabularnewline
4 & $G_{Na}$ & $mS/cm^{2}$ & $120$ & $88.8$ & $0.0$ & $150.0$\tabularnewline
5 & $G_{K}$ & $mS/cm^{2}$ & $100$ & $48.1$ & $0.0$ & $150.0$\tabularnewline
6 & $G_{K(Ca),S}$ & $mS/cm^{2}$ & $5.0$ & $3.2$ & $0.0$ & $20.0$\tabularnewline
7 & $G_{CaN,S}$ & $mS/cm^{2}$ & $14.0$ & $0.0$ & $0.0$ & $20.0$\tabularnewline
8 & $G_{K(Ca),D}$ & $mS/cm^{2}$ & $1.1$ & $0.72$ & $0.0$ & $5.0$\tabularnewline
9 & $G_{CaN,D}$ & $mS/cm^{2}$ & $0.3$ & $0.64$ & $0.0$ & $1.0$\tabularnewline
10 & $G_{CaL}$ & $mS/cm^{2}$ & $0.33$ & $0.2$ & $0.0$ & $1.0$\tabularnewline
11 & $V_{H,m_{Na}}$ & $mV$ & -35.0  & $-29.7$ & -60.0 \textbf{(-45.0)} & $-20.0$ \textbf{(-25.0)}\tabularnewline
12 & $V_{H,h_{Na}}$ & $mV$ & $-55.0$ & $-48.5$ & -60.0 \textbf{(-65.0)} & $-20.0$ \textbf{(-45.0)}\tabularnewline
13 & $V_{H,m_{K}}$ & $mV$ & $-28.0$ & $-24.1$ & -60.0 \textbf{(-40.0)} & $-20.0$ \textbf{(-20.0)}\tabularnewline
14 & $V_{H,m_{CaN}}$ & $mV$ & $-30.0$ & $-33.2$ & -60.0 \textbf{(-40.0)} & $-20.0$ \textbf{(-20.0)}\tabularnewline
15 & $V_{H,h_{CaN}}$ & $mV$ & $-45.0$ & $-41.4$ & -60.0 \textbf{(-55.0)} & $-20.0$ \textbf{(-35.0)}\tabularnewline
16 & $V_{H,m_{CaL}}$ & $mV$ & $-40.0$ & $-45.4$ & -60.0 \textbf{(-50.0)} & $-20.0$ \textbf{(-30.0)}\tabularnewline
17 & $V_{S,m_{Na}}$ & $mV$ & $7.8$ & $8.9$ & $5.0$ \textbf{(5.0)} & $25.0$ \textbf{(10.0)}\tabularnewline
18 & $V_{S,h_{Na}}$ & $mV$ & $-7.0$ & $-12.7$ & $-25.0$ \textbf{(-10.0)} & $-5.0$ \textbf{(-5.0)}\tabularnewline
19 & $V_{S,m_{K}}$ & $mV$ & $15.0$ & $21.7$ & $5.0$ \textbf{(10.0)} & $25.0$ \textbf{(20.0)}\tabularnewline
20 & $V_{S,m_{CaN}}$ & $mV$ & $5.0$ & $23.0$ & $3.0$ \textbf{(3.0)} & $23.0$ \textbf{(8.0)}\tabularnewline
21 & $V_{S,h_{CaN}}$ & $mV$ & $-5.$0  & $-5.4$ & $-23.0$ \textbf{(-8.0)} & $-3.0$ \textbf{(-3.0)}\tabularnewline
22 & $V_{S,m_{CaL}}$ & $mV$ & $7.0$ & $19.8$ & $5.0$ \textbf{(5.0)} & $25.0$ \textbf{(10.0)}\tabularnewline
23 & $\tau_{min,m_{K}}$ & $ms$ & $0.65$ & $0.2$ & $0.01$ & $1.0$\tabularnewline
24 & $\tau_{max,h_{Na}}$ & $ms$ & $30.3$ & $11.6$ & $0.01$ & $70.0$\tabularnewline
25 & $\tau_{max,m_{K}}$ & $ms$ & $6.3$ & $7.3$ & $0.01$ & $10.0$\tabularnewline
26 & $\delta_{h_{Na}}$ & - & $0.6$ & $0.2$ & $0.0$ \textbf{(0.5)} & $1.0$ \textbf{(1.0)}\tabularnewline
27 & $\delta_{m_{K}}$ & - & $0.7$ & $0.7$ & $0.0$ \textbf{(0.5)} & $1.0$ \textbf{(1.0)}\tabularnewline
28 & $\tau_{m_{CaN}}$ & $ms$ & $4.0$ & $9.8$ & $0.01$ & $10.0$\tabularnewline
29 & $\tau_{h_{CaN}}$ & $ms$ & $40.0$ & $17.0$ & $0.01$ & $70.0$\tabularnewline
30 & $\tau_{m_{CaL}}$ & $ms$ & $40.0$ & $48.1$ & $0.01$ & $70.0$\tabularnewline
\hline 
\end{tabular}

$^{1}$These parameter values were estimated when we used the broad
prior intervals (see Fig. 11Ai)

$^{2}$Values in bold indicate the narrow prior intervals we used
for generating Figs. 11Aii, 11B, 11C (and Supplementary Figs. S4 and
S5) 
\end{table}

\begin{table}[!ht]
\caption{\textbf{Estimated mean values and prior limits used during smoothing
for all parameters in the B4 model} }

\begin{tabular}{|c|l|l|r|r|r|}
\hline 
\# & Parameter & Unit & Estimated Mean Value$^{1}$ & Lower Bound & Upper Bound\tabularnewline
\hline 
1 & $G_{Na}$ & $mS/cm^{2}$ & $24.9$ & $0.0$ & $60.0$\tabularnewline
2 & $G_{K}$ & $mS/cm^{2}$ & $21.5$ & $0.0$ & $60.0$\tabularnewline
3 & $G_{A}$ & $mS/cm^{2}$ & $23.3$ & $0.0$ & $60.0$\tabularnewline
4 & $V_{H,m_{Na}}$ & $mV$ & $-25.1$ & $-70.0$ \textbf{(-40.0)} & $0.0$ \textbf{(-20.0)}\tabularnewline
5 & $V_{H,h_{Na}}$ & $mV$ & $-24.1$ & $-70.0$ \textbf{(-40.0)} & $0.0$ \textbf{(-20.0)}\tabularnewline
6 & $V_{H,m_{K}}$ & $mV$ & $-23.1$ & $-70.0$ \textbf{(-40.0)} & $0.0$ \textbf{(-20.0)}\tabularnewline
7 & $V_{H,m_{A}}$ & $mV$ & $-10.2$ & $-70.0$ \textbf{(-20.0)} & $0.0$ \textbf{(0.0)}\tabularnewline
8 & $V_{H,h_{A}}$ & $mV$ & $-53.6$ & $-70.0$ \textbf{(-70.0)} & $0.0$ \textbf{(-40.0)}\tabularnewline
9 & $V_{S,m_{Na}}$ & $mV$ & $6.6$ & $5.0$ \textbf{(5.0)} & $25.0$ \textbf{(10.0)}\tabularnewline
10 & $V_{S,h_{Na}}$ & $mV$ & $-6.5$ & $-25.0$ \textbf{(-10.0)} & $-5.0$ \textbf{(-5.0)}\tabularnewline
11 & $V_{S,m_{K}}$ & $mV$ & $11.0$ & $5.0$ \textbf{(10.0)} & $25.0$ \textbf{(15.0)}\tabularnewline
12 & $V_{S,m_{A}}$ & $mV$ & $6.8$ & $5.0$ \textbf{(5.0)} & $25.0$ \textbf{(10.0)}\tabularnewline
13 & $V_{S,h_{A}}$ & $mV$ & $-20.1$ & $-25.0$ \textbf{(-25.0)} & $-5.0$ \textbf{(-15.0)}\tabularnewline
14 & $\tau_{max,h_{Na}}$ & $ms$ & $22.9$ & $0.01$ \textbf{(15.0)} & $60.0$ \textbf{(25.0)}\tabularnewline
15 & $\tau_{max,m_{K}}$ & $ms$ & $32.0$ & $0.01$ \textbf{(25.0)} & $60.0$ \textbf{(35.0)}\tabularnewline
16 & $\tau_{max,m_{A}}$ & $ms$ & $29.5$ & $0.01$ \textbf{(25.0)} & $60.0$ \textbf{(35.0)}\tabularnewline
17 & $\tau_{max,h_{A}}$ & $ms$ & $49.9$ & $0.01$ \textbf{(35.0)} & $60.0$ \textbf{(60.0)}\tabularnewline
\hline 
\end{tabular}

$^{1}$These parameter values were estimated when we used the narrow
prior intervals (in bold; see Fig. 12)

$^{2}$The parameter posteriors estimated when we used the broad prior
intervals are illustrated in Supplementary Fig. S7 
\end{table}

\end{document}


\begin{flushleft}
{\Large
\textbf{A Self-Organizing State-Space-Model Approach for Parameter Estimation in Hodgkin-Huxley-Type Models of Single Neurons}
}
\\
Dimitrios V. Vavoulis$^{1,\ast}$, 
Volko A. Straub$^{2}$, 
John A. D. Aston$^{3}$,
Jianfeng Feng$^{1,*}$
\\
\bf{1} Dept. of Computer Science, University of Warwick, Coventry, UK
\\
\bf{2} Dept. of Cell Physiology and Pharmacology, University of Leicester, Leicester, UK
\\
\bf{3} Dept. of Statistics, University of Warwick, Coventry, UK
\\
$\ast$ E-mail: Dimitris.Vavoulis@dcs.warwick.ac.uk, Jianfeng.Feng@warwick.ac.uk
\end{flushleft}

\section*{\noindent Supplementary Material}

There are several ways to introduce noise in the Hodgkin-Huxley-type
neuron models as the ones we examined in this paper\cite{Feng2004,Goldwyn:2011fk}.
A quite common approach is to add a white noise term in the right-hand
side of the current conservation equation (which describes the evolution
of the membrane potential in time), as seen for example in Eq. 31
in the main text. This ``noisy current'' aims to approximate the
effect of a number of factors, such as the stochastic opening and
shutting of transmembrane ion channels or the random bombardment of
the neuron with synaptic input, and its major advantage is its simplicity.
This is the approach we followed in this study. Since a major source
of noise is the random fluctuations in the total conductance within
a population of ion channels, it is reasonable to assume that similar
(possibly, state-dependent) noise terms should be included in the
dynamic equations describing the time evolution of the activation
and inactivation gating variables (Eq. 32). For a single compartment
model (as in Eqs. 31 and 32 in the main text), we can write:
\begin{equation}
dV=\frac{I_{ext}-G_{L}(V-E_{L})-G_{Na}m^{3}h(V-E_{Na})-G_{K}n^{4}(V-E_{K})}{C_{m}}dt-\frac{1}{C_{m}}dI_{syn}\label{mdl:HH1}
\end{equation}
\begin{equation}
dx=\left(a_{x}(1-x)-b_{x}x\right)dt+\sigma_{X}\sqrt{a_{x}(1-x)+b_{x}x}dW_{x}\label{mdl:HH2}
\end{equation}
where $x\in\{m,h,n\}$, $X\in\{Na,K\}$ and $\sigma_{X}=\left(\sqrt{N_{X}}\right)^{-1}$
with $N_{X}$ being the total number of sodium or potassium channels
in the model. $a_{x}$ and $b_{x}$ are functions of voltage, as shown
below:
\begin{eqnarray*}
a_{m}=0.1\frac{V+40}{1-\exp\left(-\frac{V+40}{10}\right)} & \qquad,\qquad & b_{m}=4\exp\left(-\frac{V+65}{18}\right)\\
a_{h}=0.07\exp\left(-\frac{V+65}{20}\right) & \qquad,\qquad & b_{h}=\frac{1}{1+\exp\left(-\frac{V+35}{10}\right)}\\
a_{n}=0.01\frac{V+55}{1-\exp\left(-\frac{V+55}{10}\right)} & \qquad,\qquad & b_{n}=0.125\exp\left(-\frac{V+65}{80}\right)
\end{eqnarray*}
Notice that the noise terms in Eq. \ref{mdl:HH2} depend on both the
voltage and the gating variables. Also notice that, in Eq. \ref{mdl:HH1},
$I_{syn}$ is the sum of the excitatory and inhibitory synaptic input
the neuron receives. For an infinitesimal change in this current,
we can write:
\begin{equation}
dI_{syn}=\gamma_{E}(V-E_{E})dP_{E}+\gamma_{I}(V-E_{I})dP_{I}
\end{equation}
where $dP_{E}$ and $dP_{I}$ are Poisson processes, which model the
random arrival of presynaptic excitatory and inhibitory spikes at
firing rates $\lambda_{E}$ and $\lambda_{I}$, respectively. $\gamma_{E}$
and $\gamma_{I}$ are unitary increases in the synaptic conductance
and $E_{E}$ and $E_{I}$ are the reversal potentials of the excitatory
and inhibitory synaptic currents, respectively. Assuming that the
neuron receives a high-frequency barrage of presynaptic spikes, it
is common to re-write the above expression for synaptic current using
the diffusion approximation\cite{Hanson1983}:
\begin{equation}
dI_{syn}=\left(\gamma_{E}\lambda_{E}(V-E_{E})+\gamma_{I}\lambda_{I}(V-E_{I})\right)dt+\sqrt{\lambda_{E}\gamma_{E}^{2}(V-E_{E})^{2}+\lambda_{I}\gamma_{I}^{2}(V-E_{I})^{2}}dW_{syn}\label{mdl:HH3}
\end{equation}
Notice that we have assumed that changes in the total synaptic current
are instantaneous. This is just an approximation, since changes in
synaptic conductances have characteristic rise and decay relaxation
times (see, for example, \cite{Destexhe:2001uq,Richardson:2005fk}).
Observation noise was as in Eq. 7 in the main text with $\sigma_{y}=1mV$. 

In Eq. \ref{mdl:HH1}, the membrane capacitance, maximal conductances
and reversal potentials were as follows: $C_{m}=1nF/cm^{2}$, $G_{L}=0.3mS/cm^{2}$,
$G_{Na}=120mS/cm^{2}$, $G_{K}=36mS/cm^{2}$, $E_{L}=-54.4mV$, $E_{Na}=55mV$,
$E_{K}=-77mV$, $E_{E}=0mV$ and $E_{I}=-75mV$. In Eq. \ref{mdl:HH2},
$\sigma_{Na}=0.04$ and $\sigma_{K}=0.02$. Unitary synaptic conductances
and presynaptic firing rates in Eq. \ref{mdl:HH3} were: $\gamma_{E}=1mS/cm^{2}$,
$\gamma_{I}=1mS/cm^{2}$, $\lambda_{E}=0.03ms^{-1}$ and $\lambda_{I}=0.01ms^{-1}$.
With these parameters, the model in Eqs. \ref{mdl:HH1}, \ref{mdl:HH2}
and \ref{mdl:HH3} was active in the absence of any external input
$I_{ext}$. Given a recording of this activity, the fixed lag-smoother
can be used for retrieving the hidden states of the model and various
parameters that control channel and synaptic noise ($\sigma_{Na}$,
$\sigma_{K}$, $\lambda_{E}$ and $\lambda_{I}$), as shown in Figs.
S1 and S2. This simulation experiment demonstrates the applicability
of the algorithm, when more complex noise models are considered. 

\bibliography{supplementary}

\clearpage

\section*{Supplementary Figures}

\begin{figure}[!ht]
\includegraphics[width=4in]{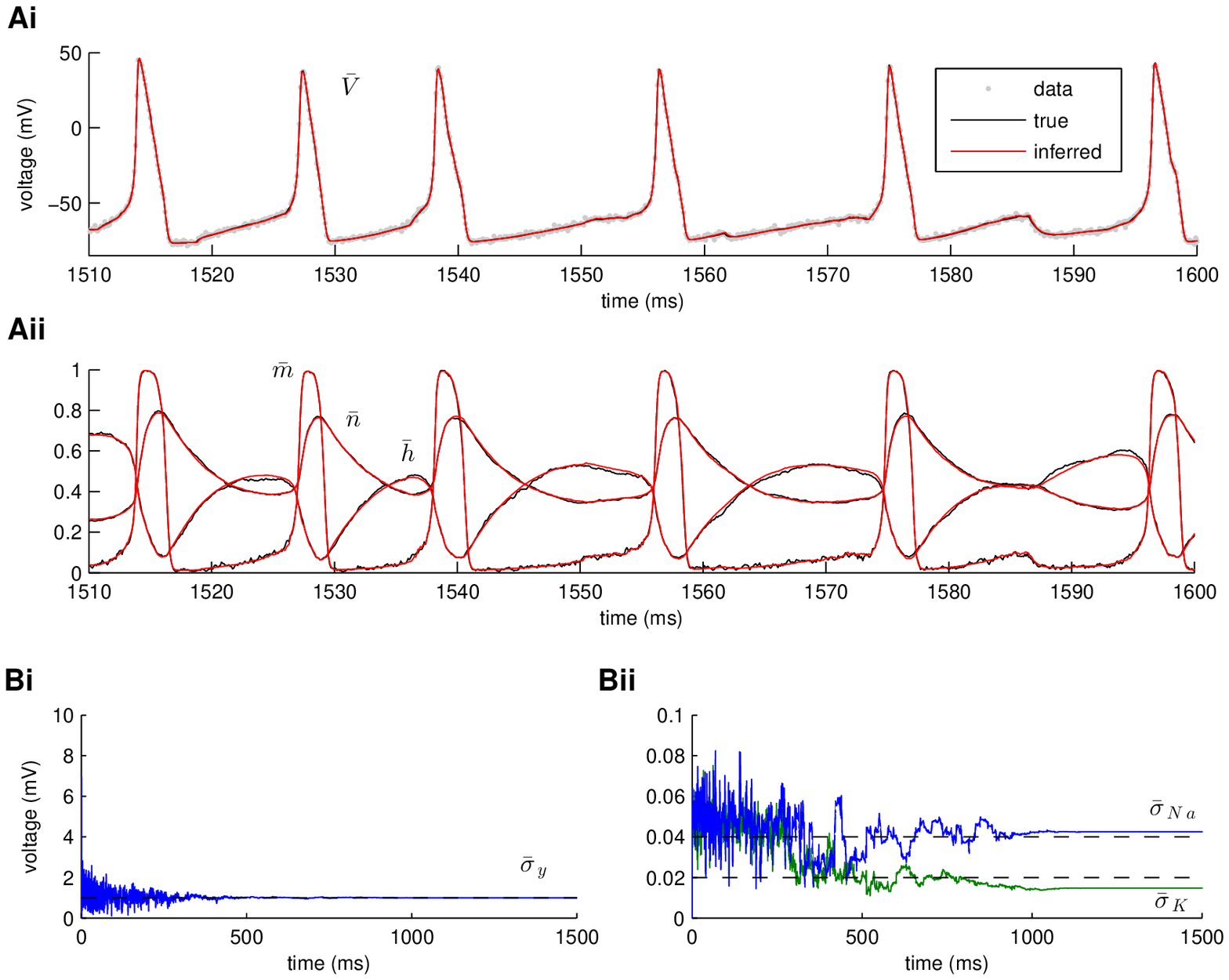}
\caption{\textbf{Simultaneous estimation of hidden states and channel noise
in a stochastic single-compartment model}. Estimation was based on
a simulated $1s$-long recording of membrane potential generated by
Eqs. \ref{mdl:HH1}, \ref{mdl:HH2} and \ref{mdl:HH3}. For clarity,
only $90ms$ of activity are shown in Figs. Ai,ii. Notice that in
these simulations, we assumed the absence of synaptic input (i.e.
$\gamma_{E}=\gamma_{I}=0mS/cm^{2}$). Activity in the model neuron
was driven by a random sequence of current steps $I_{ext}$ with amplitude
between $-5\mu A/cm^{2}$ and $20\mu A/cm^{2}$ and duration up to
$20ms$. (\textbf{A}) Simultaneous inference of the observed membrane
potential (Ai) and the hidden activation ($m$, $n$) and inactivation
($n$) gating variables for the sodium and potassium currents. (\textbf{B})
Inference of the standard deviation of the observation noise $\sigma_{y}$
(Bi) and the parameters $\sigma_{Na}$ and $\sigma_{K}$, which control
the variance of the sodium and potassium channel noise (Bii). Estimates
converged to their final values after approximately $1000ms$. The
dashed lines indicate the true values of these parameters. The y-axes
in Bi,ii indicate the width of the prior intervals imposed on the
corresponding parameters. Simulation parameters were: $L=100$, $N=800$
and $a=b=c=0.01$. The prior interval for the scaling factors $s_{k}^{(j)}$
was $[0,2]$. }
\end{figure}

\begin{figure}[!ht]
\includegraphics[width=4in]{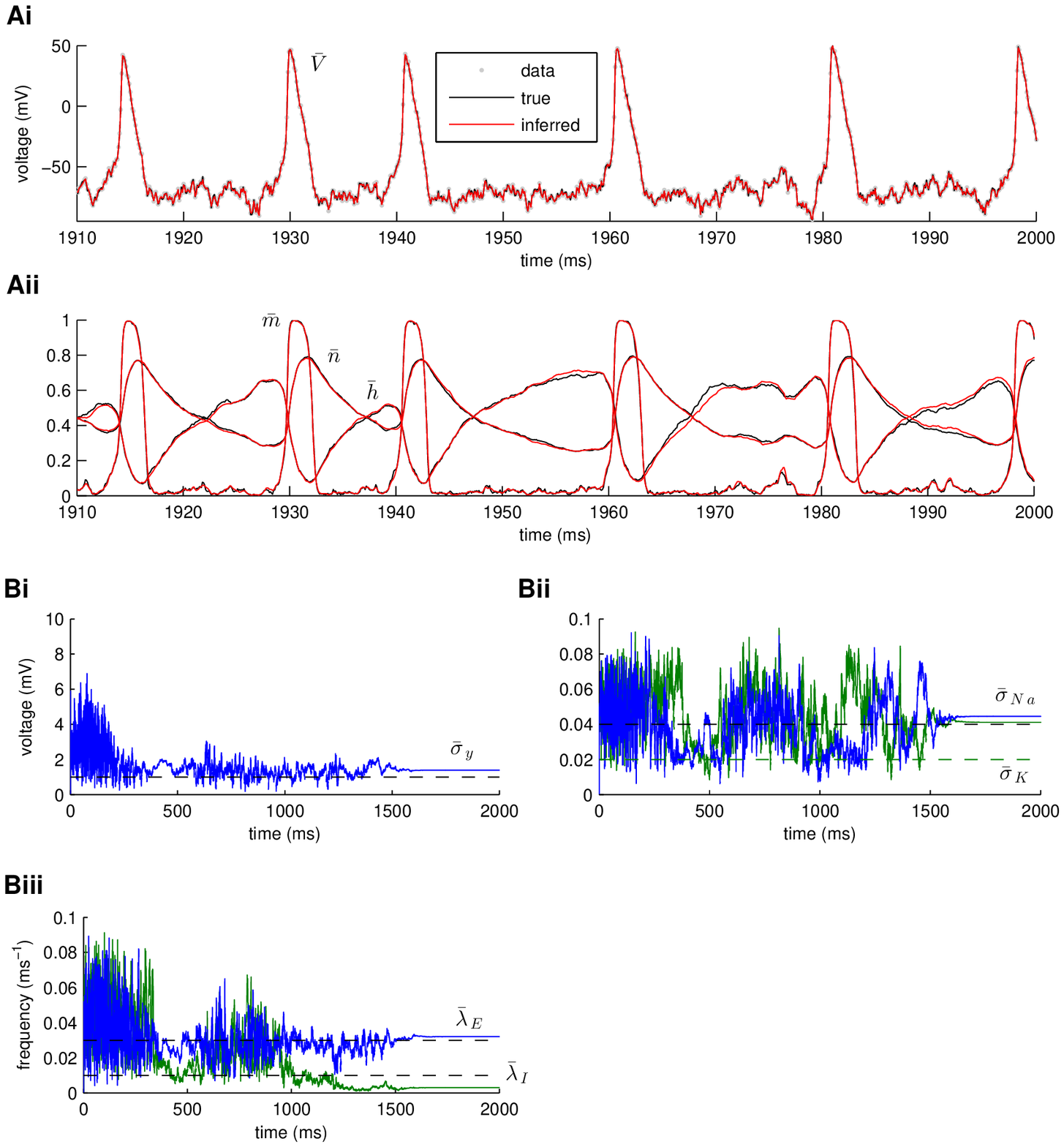}
\caption{\textbf{Simultaneous estimation of hidden states, channel noise and
presynaptic firing rates in a stochastic single-compartment model}.
Estimation was based on a simulated $2s$-long recording of membrane
potential generated by Eqs. \ref{mdl:HH1}, \ref{mdl:HH2} and \ref{mdl:HH3}
with $I_{ext}=0\mu A/cm^{2}$. For clarity, only $90ms$ of activity
are shown in Figs. Ai,ii. (\textbf{A}) Simultaneous inference of the
observed membrane potential (Ai) and the hidden activation ($m$,
$n$) and inactivation ($n$) gating variables for the sodium and
potassium currents (Aii). (\textbf{B}) Inference of the standard deviation
of the observation noise $\sigma_{y}$ (Bi), parameters $\sigma_{Na}$
and $\sigma_{K}$, which control the variance of the sodium and potassium
channel noise (Bii) and the presynaptic firing rates $\lambda_{E}$
and $\lambda_{I}$ (Biii). Estimates converged to their final values
after approximately $2s$ of activity. The dashed lines indicate the
true values of the parameters. The y-axes in B and C indicate the
width of the prior intervals imposed on the corresponding parameters.
Discrepancies from the true values in B are due to the overlapping
effects of different parameters controlling observation, channel and
synaptic noise. Simulation parameters were: $L=100$, $N=1000$ and
$a=b=c=0.01$. The prior interval for the scaling factors $s_{k}^{(j)}$
was $[0,2]$. }
\end{figure}

\begin{figure}[!ht]
\includegraphics[width=4in]{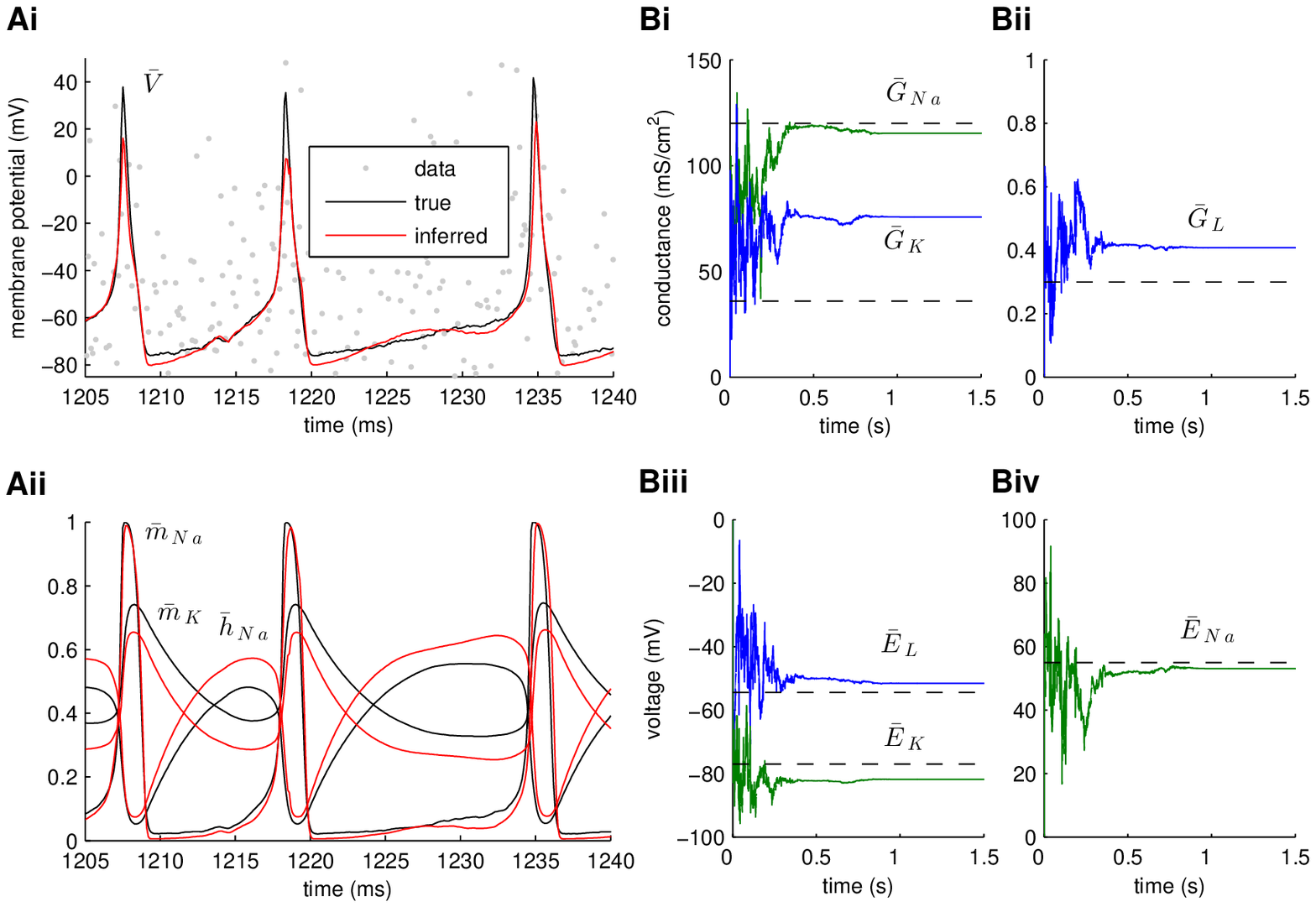}
\caption{\textbf{Simultaneous inference of hidden states and unknown parameters
in the single compartment model (see main text) at high levels of
observation noise.} This figure corresponds to Fig. 7Dii in the main
text for $\sigma_{y}=50mV$. (\textbf{A}) Inferred membrane potential
(Ai) and unobserved gating variables (Aii). (\textbf{B}) Examples
of simultaneously inferred parameters, such as maximal conductances
(Bi,ii) and reversal potentials (Biii,iv). The y-axes in Bi-iv indicate
the prior intervals of the corresponding parameters. Simulation details
are as in Fig. 7 in the main text. }
\end{figure}

\begin{figure}[!ht]
\includegraphics[width=4in]{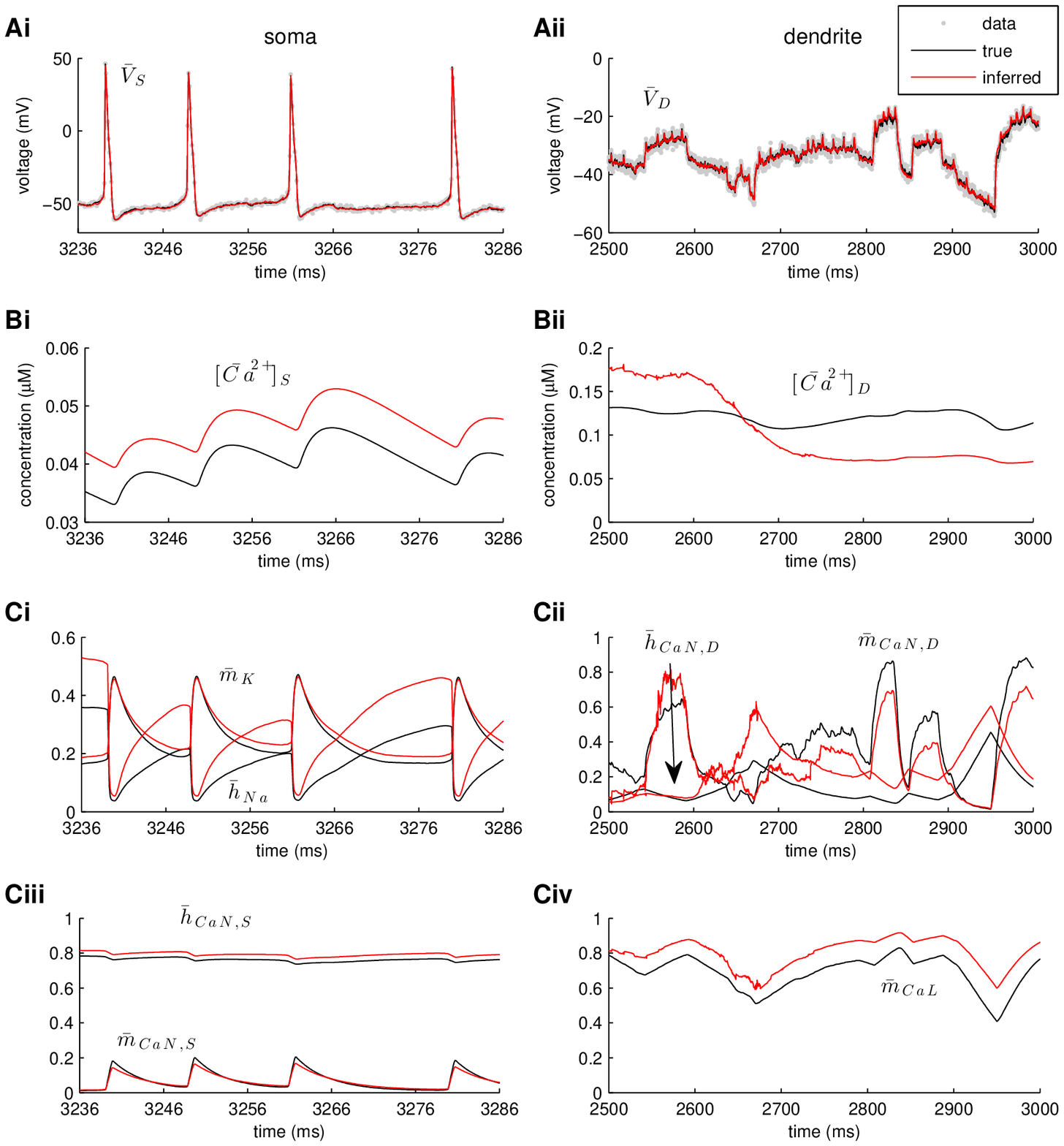}
\caption{\textbf{Simultaneous inference of hidden states in the two-compartment
model (see main text) at low levels of observation noise. }This figure
corresponds to Figs. 11Aii for $\sigma_{y}=1mV$, 11B and 11C in the
main text. (\textbf{A}) Inference of the membrane potential at the
soma (Ai) and the dendritic compartment (Aii). (\textbf{B}) Inference
of the unobserved concentration of intracellular calcium at the soma
(Bi) and the dendritic compartment (Bii). (\textbf{C}) Inference of
the unobserved gating variables for the sodium and potassium currents
at the soma (Ci), the N-type calcium current at the soma (Ciii), the
N-type calcium current at the dendritic compartment (Cii) and the
L-type calcium current at the dendritic compartment (Civ). Simulation
details are as in Fig. 11 in the main text. }
\end{figure}

\begin{figure}[!ht]
\includegraphics[width=4in]{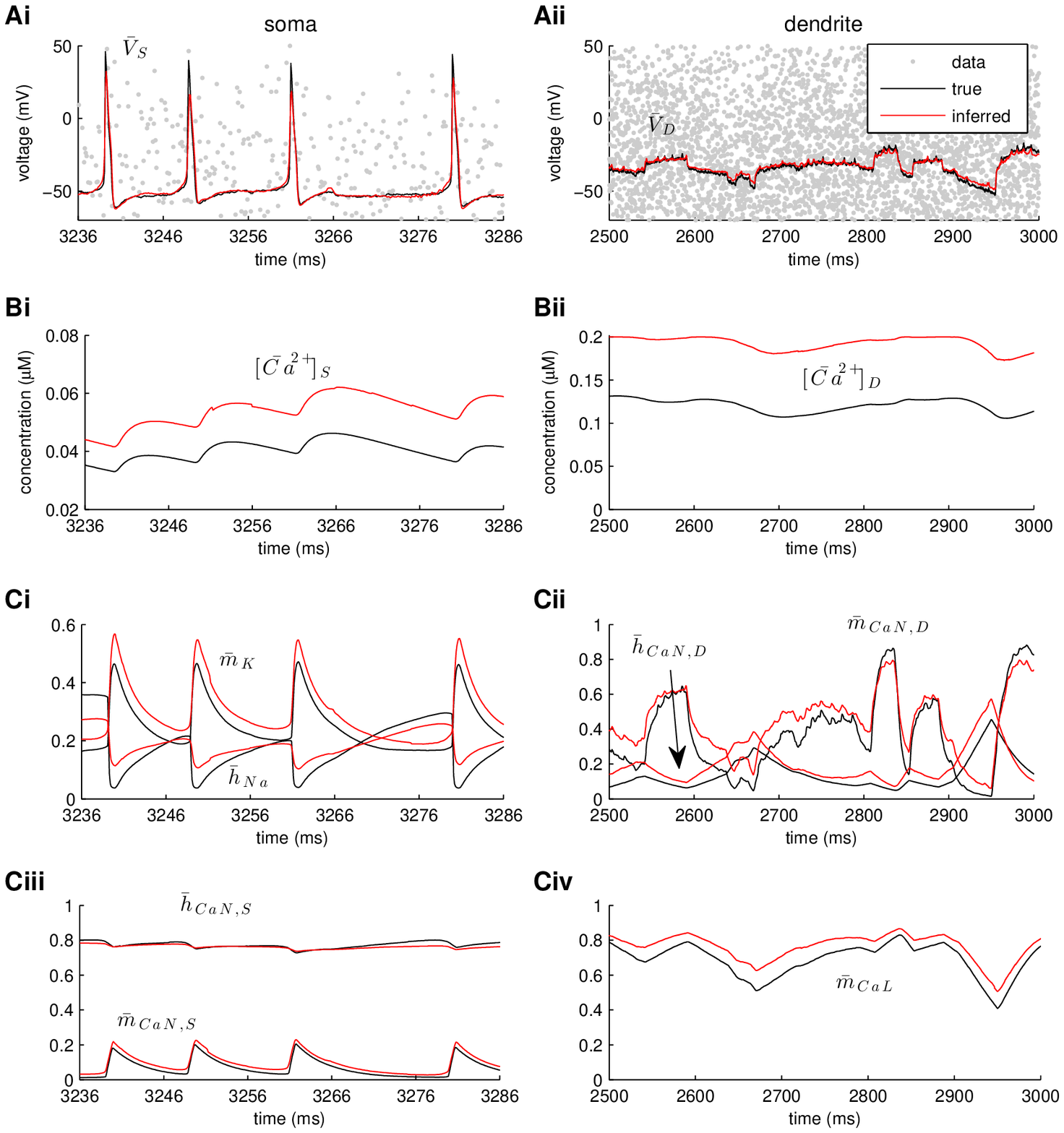}
\caption{\textbf{Simultaneous inference of hidden states in the two-compartment
model (see main text) at high levels of observation noise. }This figure
corresponds to Fig. 11Aii for $\sigma_{y}=50mV$. (\textbf{A}) Inference
of the membrane potential at the soma (Ai) and the dendritic compartment
(Aii). (\textbf{B}) Inference of the unobserved concentration of intracellular
calcium at the soma (Bi) and the dendritic compartment (Bii). (\textbf{C})
Inference of the unobserved gating variables for the sodium and potassium
currents at the soma (Ci), the N-type calcium current at the soma
(Ciii), the N-type calcium current at the dendritic compartment (Cii)
and the L-type calcium current at the dendritic compartment (Civ).
Simulation details are as in Fig. 11 in the main text. }
\end{figure}

\begin{figure}[!ht]
\includegraphics[width=4in]{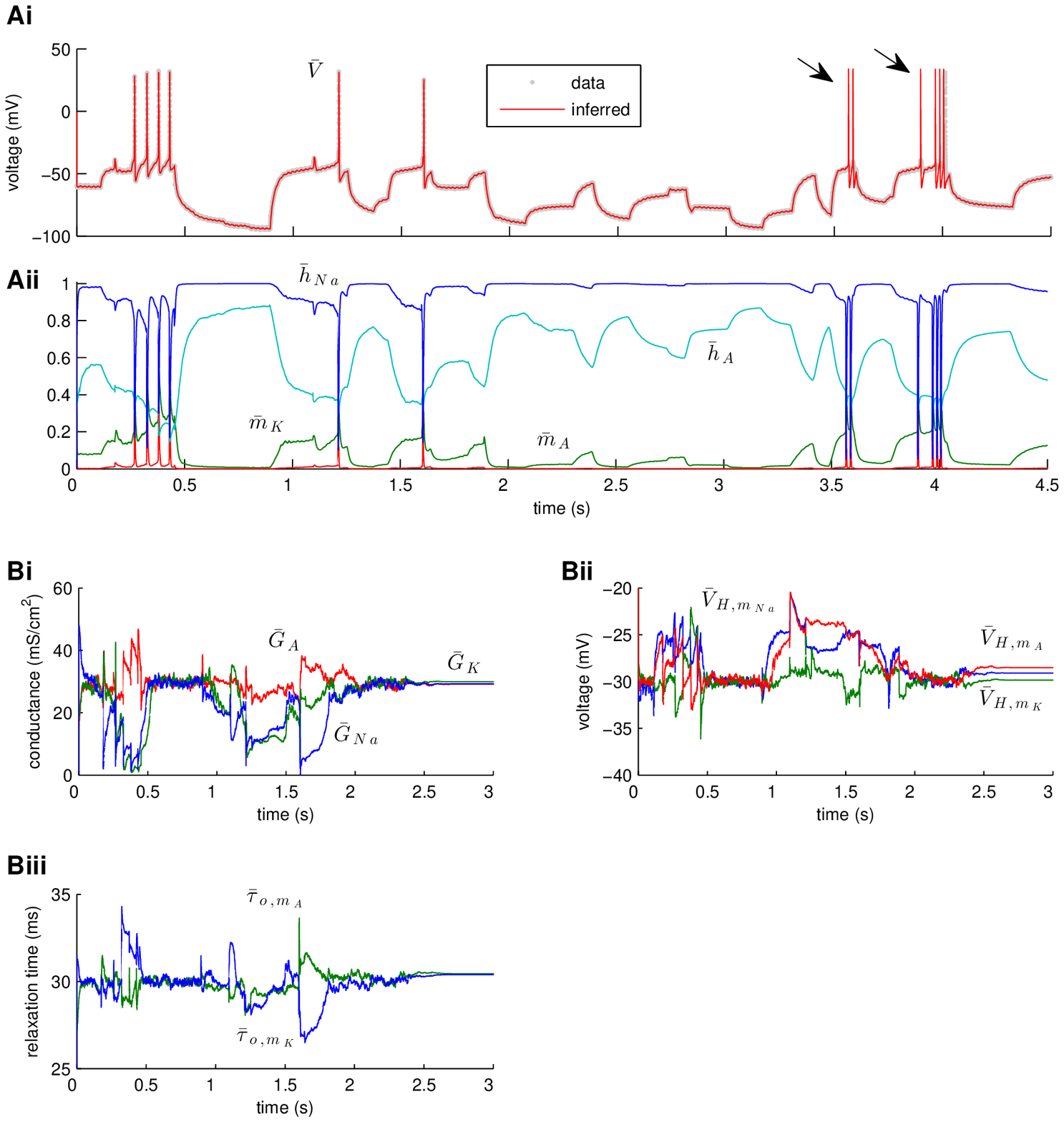}
\caption{\textbf{Inference in the B4 model using a single recording of the
membrane potential. }A single $4.5s$-long recording of B4 activity
induced by injecting a sequence of random current steps in the neuron
was used during smoothing. Random current amplitude was between $-4nA$
and $+4nA$ and random step duration was between $1ms$ and $256ms$.
(\textbf{A}) Inference of the membrane potential (Ai) and the unobserved
gating variables for the sodium and potassium currents in the model
(Aii). (\textbf{B}) Examples of simultaneously inferred model parameters:
maximal conductances of all currents (Bi), half steady-state activation
voltages for all currents (Bii) and maximal relaxation times for the
activation of the potassium currents in the model (Biii). Notice that
in all cases the parameter estimates converge exactly to the middle
of their prior intervals (indicated by the y-axes in Bi-iii). This
convergence takes place while the algorithm processes the ``inactive''
region of the data (approximately, from second $2$ to second 3 in
Ai). Based on these converged estimates, the model incorrectly emits
spikes later during smoothing (see arrows in Ai), indicating that
the estimated parameters are not optimal for smoothing during the
whole duration of experimental data. Simulation parameters were as
follows: $L=100$, $N=2300$ and $a=b=c=0.01$. The prior interval
for the scaling factors $s_{k}^{(j)}$ was $[0.2,0.5]$. For the $17$
free parameters in the model, we used the narrow prior intervals in
Table 3.}
\end{figure}

\begin{figure}[!ht]
\includegraphics[width=4in]{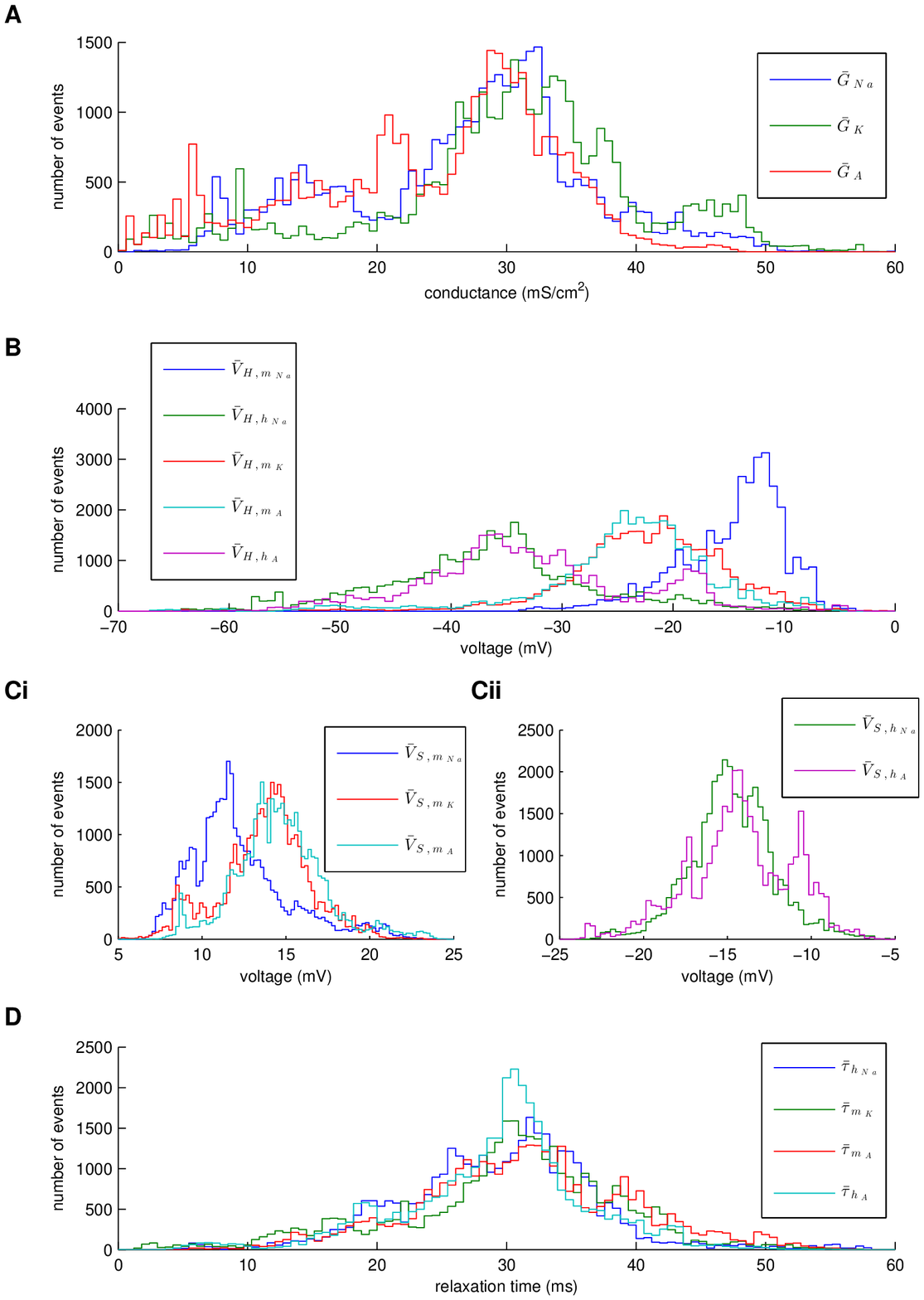}
\caption{\textbf{Inferred posterior distributions of all unknown parameters
in the B4 model using the broad prior intervals in Table 3.} Inference
was based on simultaneously smoothing four $3.5s$-long voltage recordings
from the B4 neuron as in Fig. 12A in the main text. As in that case,
data smoothing was accomplished with very high fidelity, as illustrated
in Fig12A. (\textbf{A}) Inferred maximal conductances. (\textbf{B})
Inferred half steady-state activation and inactivation voltages. (\textbf{C})
Inferred activation (Ci) and inactivation (Cii) voltage sensitivities
(parameters $V_{S,x_{i}}$ in the model). (\textbf{D}) Activation
and inactivation relaxation times. The x-axes in all plots indicate
the prior parameter intervals we used (Table 3). Notice that most
posteriors are very broad (covering a large portion of the prior interval)
and not unimodal. Simulation parameters were as described in Fig.
12 of the main text. }
\end{figure}